\documentclass[12pt,preprint]{aastex}

\shorttitle{Central Stars of Bipolar PNe}
\shortauthors{Lee et al.}

\begin{document}

\title{High resolution spectra of bright central stars of bipolar
planetary nebulae, and the question of magnetic shaping\footnote{Based
on observations made with ESO Telescopes at the La Silla Observatories
under program ID 70.D-0339} }

\author{Ting-Hui Lee and Letizia Stanghellini\altaffilmark{*}}
\affil{National Optical Astronomy Observatory, 950 North Cherry
  Avenue, Tucson, AZ 85719}
\email{thlee@noao.edu; lstanghellini@noao.edu}

\and

\author{Lilia Ferrario and Dayal Wickramasinghe}
\affil{Department of Mathematics, The Australian National University,
  Canberra, ACT 0200, Australia}
\email{lilia@maths.anu.edu.au; dayal@maths.anu.edu.au}

\altaffiltext{*}{On leave from INAF-Bologna Observatory.}

\begin{abstract}

  We present ESO NTT high resolution echelle spectroscopy of the
  central stars (CSs) of eight southern bipolar planetary nebulae
  (PNe) selected for their asymmetry. Our aim was to determine or
  place limits on the magnetic fields of the CSs of these nebulae, and
  hence to explore the role played by magnetic fields in nebular
  morphology and PN shaping. If magnetic fields do play a role, we
  expect these CSs to have fields in the range $10^2 - 10^7$~G from
  magnetic flux conservation on the reasonable assumption that they
  must evolve into the high field magnetic white dwarfs.  We were able
  to place an upper limit of $\approx 20,000$ G to the magnetic fields
  of the central stars of He 2-64 and MyCn 18. The spectrum of He~2-64
  also shows a P-Cygni profile in \ion{He}{1} $\lambda$5876 and
  $\lambda$6678, corresponding to an expanding photosphere with
  velocity $\sim$ 100 km s$^{-1}$.  The detection of helium absorption
  lines in the spectrum of He~2-36 confirms the existence of a hot
  stellar component.  We did not reach the necessary line detection
  for magnetic field analysis in the remaining objects.  Overall, our
  results indicate that if magnetic fields are responsible for shaping
  bipolar planetary nebulae, these are not required to be greater than
  a few tens of kilogauss.

\end{abstract}


\section{Introduction}

  Planetary Nebulae (PNe) and their central stars (CSs) are the
  transition phase between asymptotic giant branch (AGB) stars and
  white dwarfs (WDs) for main-sequence masses between 1 and
  8~M$_{\odot}$.  PNe are found in a wide range of morphologies, but
  the majority of shapes can be classified into three types: round,
  elliptical, and bipolar (or multipolar) PNe \citep{schwarz93,
  manchado96}.  Bipolar and multipolar PNe are characterized by a
  waist, and show one or more sets of lobes.  Several mechanisms have
  been proposed for shaping the bipolar structures, including common
  envelope evolution \citep[e.g.,][]{bodenheimer84}, magnetic fields
  associated with the central stars \citep[e.g.,][]{pascoli92}, and
  stellar rotation \citep[e.g.,][]{ignace96}.

  Magnetic fields have been an attractive hypothesis for shaping PNe
  into bipolar morphology for single stars.  From a theoretical point
  of view, a toroidal magnetic field embedded in a spherically
  expanding fast wind is able to reproduce the majority of the PN
  morphologies \citep{garcia99}.  As explored by \cite{garcia05}, if the
  magnetic field is strong enough to drive the AGB winds, magnetic
  pressure alone can also drive the winds into high speeds as observed
  in proto-planetary nebulae \citep{bujarrabal01}.  Other models
  involved the magneto-centrifugal processes have also been applied to
  PNe and shown that such processes can efficiently power and
  collimate the outflows \citep{blackman01}.

  Interestingly, magnetic fields have been detected in the progenitors
  and the progeny of PNe, namely, around proto-planetary nebulae
  (pPNe), and in WDs.  The observations of magnetic fields in pPNe are
  mostly from SiO, OH, and H$_{2}$O maser emissions
  \citep[e.g.,][]{miranda01, bains04}.  Recently, \cite{vlemmings06}
  have found direct evidence that the magnetic field is collimating
  the jet from the AGB star W43A.  They measured the polarization of
  water vapor masers that trace the precessing jet emanating from the
  star, and concluded that the magnetic field and jet characteristics
  of W43A support the shaping mechanism of PNe by magnetically
  collimated jets from evolved stars.  

  Concerning the progeny of PNe, it has been known that a fraction of
  WDs have detectable magnetic fields.  Since the discovery of the
  first isolated magnetic WD (Grw~+70$^{\circ}$8047 was discovered by
  \citealt{minkowski38}, though its peculiar spectrum was not
  recognized to be caused by a strong magnetic field until much later
  by \citealt{angel85} and \citealt{greenstein85}), the number of
  detected magnetic white dwarfs (MWDs) have grown steadily \citep[and
  references therein]{wickramasinghe00}.  Recent studies show that
  least 10\% of WDs have magnetic fields at the level of $\sim$ 2~MG
  and larger, and this fraction increases if low-field objects are
  included \citep{liebert03}.  Since PNe and their CSs are the
  evolutionary link between pPNe and WDs, at least a fraction of them
  should also contain magnetic fields.  However, such observational
  evidence has been scarce.  Very recently, some hints of the
  existence of magnetic fields in PNe have been given by means of
  spectropolarimetry \citep{jordan05}.  Among the four CSs of PNe
  observed by \cite{jordan05}, they found possible signature of
  magnetic fields in two CSs.

  Isolated MWDs with fields $\gtrsim 10^6 $ G show a mean mass of
  $\sim$ 0.93 M$_{\odot}$, compared to the main peak of the mass
  distribution of $\sim$ 0.57 M$_{\odot}$ for the nonmagnetic white
  dwarfs \citep[and references therein]{wickramasinghe00}.  This is
  interesting, since the statistical studies of PNe have shown that
  bipolar PNe may have more massive progenitors: Observations from a
  large sample of Galactic PNe have shown that bipolar PNe are closer
  to the Galactic plane and have more massive CSs than round and
  elliptical PNe.  Furthermore, bipolar PNe are carbon-poor and
  nitrogen-rich, consistent with a massive post-AGB stellar population
  \citep{stanghellini93, stanghellini02, manchado04}.  The opposite is
  true for round and elliptical PNe.  Recent studies of PNe in the
  Large Magellanic Cloud \citep{stanghellini00} have also provided
  more evidence that aspheric (bipolar, quadrupolar and maybe
  point-symmetric) PNe belong to a different population than round and
  elliptical PNe: they are evolved from higher mass progenitors than
  round PNe.

  The high mass distribution of isolated MWDs, together with more
  massive progenitors of bipolar PNe, seem to indicate that isolated
  MWDs could be the evolutionary product of magnetic CSs.  If this
  hypothesis is correct, then, under the condition of magnetic flux
  conservation, we should expect a CS to MWD field ratio of
\begin{equation}
 (B_{\rm CS}/B_{\rm MWD}) \sim(R_{\rm MWD}/R_{\rm CS})^2,
\end{equation}
  where {\it B} is the magnetic strength and {\it R} is the radius of
  the star.  Depending on the mass of the CS and how evolved it is,
  $R_{\rm MWD}/R_{\rm CS}$ could have a range of $\sim 0.01 - 0.1$.
  Therefore with a range of magnetic fields of $\sim 10^6 - 10^9$ G
  found in MWDs, we should expect CS magnetic fields in the range of
  $\sim 10^2 - 10^7$ G.  Thus our aim is to search for magnetic fields
  in CSs of bipolar PNe in this range in order to investigated if CSs
  of bipolar PNe can be progenitors of MWDs.

  For this study, we have obtained high resolution echelle
  spectroscopy of central stars of southern bipolar nebulae.  Our goal
  is to look for Zeeman splitting of stellar lines, caused by magnetic
  fields associated with the CSs. Our spectral resolution allows us to
  observe fields $\gtrsim 10^4$ G.

  The observations and data reduction are described respectively in
  Section \ref{sec:obs} and \ref{sec:cal}.  The stellar spectra are
  presented in Section \ref{sec:results}. Discussion of the Zeeman
  split limits and detections are in $\S$ 5, and conclusions are in
  $\S$6.

\section{Observations}
\label{sec:obs}

  Our target list includes stars at the top of the white dwarf cooling
  sequence.  The PNe hosting the targeted CSs have been previously
  observed in the major nebular narrow-band filters \citep{balick87,
  schwarz92, manchado96}, and their morphological types have been
  classified uniformly following the \cite{manchado96} scheme.  We
  selected bipolar PNe and other asymmetric PNe whose central stars
  are bright, and well separated from the nebulae.  We used these
  criteria so that the targets are easier to place on the slit, have
  higher S/N, and the stellar spectra have less nebular contribution.
  We have also checked against the list of symbiotic stars in
  \cite{corradi03} to exclude any symbiotic nebulae in our target
  list.  Our sample comprises 8 CSs which are the only stars bright
  enough to be searched for magnetic fields in the southern sky with a
  4-m class telescope.  The properties of the targets are listed in
  Table \ref{tb:source}, with column 1 and 2 giving the common name
  and the PN G name based on their galactic coordinates
  \citep{acker92}, columns 3 and 4 giving their published B and V CS
  magnitudes, when available, column 5 listing the references of the
  magnitudes, columns 6 and 7 giving their Zanstra temperatures
  derived from \ion{He}{2} in the literature (with the exception of
  energy-balance temperature for He~2-64), and column 8 listing the
  temperature references.

  The observations were done with the ESO Multi-Mode Instrument (EMMI)
  on the 3.58m New Technology Telescope (NTT) in 2003 February 4 and
  5.  We have observed our targets both with echelle spectroscopy and
  wide field imaging.  Both nights were in photometric sky conditions
  and sub-arcsecond seeing.  The filters used in wide-field imaging
  are listed in Table \ref{tb:filters} and the observing logs are
  listed in Table \ref{tb:ob-log}.  We obtained UBVR images and
  echelle spectral of 8 PNe.

  EMMI has a blue arm and a red arm.  The blue arm has one Tektronic
  CCD camera giving image size of 1024x1024.  It was operated in the
  BIMG mode to obtain U- and B-band images.  The pixel size is 24
  $\mu$m which corresponds to 0.37$''$ in the sky.  The red arm of
  EMMI has two MIT/LL CCD chips arranged in a mosaic.  The two-chip
  mosaic is read in four output mode via four separate amplifiers,
  giving slightly different values for the bias level, gain, and
  read-out noise.  Each output has image size of 1048x4096, giving
  full image size of 4096x4096 with a gap of 47 pixels in the x
  direction in between two chips.  The red arm was operated in the
  RILD mode to obtain V- and R-band images and in the REMD mode to
  obtain echelle spectra. The 2x2 binning mode was used in order to
  avoid oversampling problems, resulting a full image size of
  2048x2048 which has a pixel size of 15 $\mu$m corresponding to
  0.33$''$ in the sky.

  The echelle spectra were observed with the grating \#14 and the
  cross-dispersing grism \#5, corresponding to a wavelength range of
  4120-6730 \AA, with a gap between 5010-5060 \AA.  The dispersion in
  the 2x2 binning mode is 0.04 \AA~per pixel.  The grating \#14 gives a
  spectral resolution of R=60,000, corresponding to $\sim$ 0.1 \AA~in
  this wavelength range.

  The calibration data were taken each night in order to correct
  several effects of the CCD electronics.  The overscan region in the
  blue CCD was used to measure the electronic pedestal level of the
  CCD chip.  The red CCD is very stable and an overscan region was not
  required.  We acquired several bias frames to derive the column to
  column variations of the bias structures.  A set of 5-10 bias frames
  were taken for each CCD at the beginning, the end, and during each
  observation night with the same CCD parameters.  The dark current
  was negligible during the observations thus no dark frames were
  taken.  Finally, dome flats were obtained for each observing
  configuration.  In addition, standard star observations and
  Thorium-Argon arc spectra were taken for flux and wavelength
  calibrations.

\section{Calibration}
\label{sec:cal}

  The imaging data were calibrated following the standard techniques
  using IRAF software \citep{massey97}.  The bias frames and flats
  were combined by averaging them.  The images were then bias
  subtracted, trimmed, and flat-field corrected.  We show the R-band
  images of the targets in Figure \ref{fg:imager1} and Figure
  \ref{fg:imager2}.

  The echelle spectra were calibrated following the steps outlined in
  \cite{willmarth94}.  The target spectra, standard star spectra, flat
  fields, and wavelength calibration files were first bias subtracted
  using an average bias frame, then trimmed to remove useless parts as
  well as blue orders that are impossible to trace.  The average flats
  are normalized using the task {\tt apflatten}.  The task finds and
  traces the orders and normalizes the flat by fitting the intensity
  along the order.  The target spectra, standard star spectra, and
  wavelength calibration files were then divided by the normalized
  flat to correct for flat fielding.

  The task {\tt doecslit}, within the imred.echelle package, was used
  for spectral extraction.  The extracting orders were first defined
  and traced with the standard star HR5501 spectra.  The extracting
  aperture size is 8 pixels, which roughly corresponds to the width of
  the profile of the order at 5\% of its maximum around
  $\lambda=5500$~\AA.  We have also used an extracting aperture size
  of 4 pixels to extract the spectra in order to minimize the nebular
  contamination while studying the stellar spectra.  However, since
  the S/N get much lower as we decrease the extraction aperture, we
  only present the spectra of 8-pixel extraction and perform line
  measurements, unless otherwise noted.  The Thorium-Argon arc spectra
  were extracted for wavelength calibration.  After the wavelength of
  the entire spectral range has been defined, the observed fluxes of
  the standard star were compared against tabulated values.  For each
  order a set of bandpasses was defined.  The tabulated data were
  interpolated to the bandpasses specified.  The ratio of the observed
  flux over the bandpass to the tabulated value over the same bandpass
  is fitted by a smooth curve to relate system sensitivity to
  wavelength for each order.  This allows to derive calibration
  curves, that are then applied to produce the final flux-calibrated
  spectra.  The procedure also corrects for atmospheric extinction.

  The nebular emission line ratios were used to compare to the values
  in the literature as a test of our calibration.  We used the task
  {\tt splot} to measure the flux and of each nebular emission line.
  When applicable, we fit a Gaussian profile of the line to measure
  the flux.  When the line structure is more complex (e.g., with two
  peaks), we sum up the area to get the total flux of the line.  In
  both cases the continuum is fitted and subtracted by the task.  In
  Table \ref{tb:line_ratios}, we list the major nebular emission line
  intensities normalized to F(H$\beta$)=100 for each object in columns
  3 to 7.  Column 8 gives the extinction constant $c$ (the logarithmic
  extinction at H$\beta$) derived from the H$\alpha$ to H$\beta$
  ratio.  The central star of HDW 5 lies outside of the nebular thus
  the nebula was not included in the observation.  A comparison of our
  line intensities (versus H$\beta$) and those from the literature is
  illustrated in Figure \ref{fg:line_ratios}, which shows that our
  results generally agree well with previous observations.

\section{Results and Analysis}
\label{sec:results}

  Overall the spectra of seven objects, except HDW~5, show many
  nebular features.  He~2-25, NGC~2818, and He~2-186 also show high
  expansion velocities in the hydrogen emission lines.  In this paper
  we focus our attention to the stellar spectra, while nebular
  analysis will be presented in a future paper.  We found no evidence
  of stellar emission or absorption in three objects, He~2-25,
  He2-186, and NGC~2818.  HDW~5 has one stellar absorption line
  detected, and He~2-123 has two stellar \ion{C}{4} emission line
  detected.  Only in He~2-36, He~2-64, and MyCn~18 we detected several
  stellar lines.  We perform Gaussian fit of all stellar lines, and
  measure its flux and equivalent width.  Table \ref{tb:sline} lists
  all the lines that have been identified and measured.  Here we
  describe each object and its spectrum in detail.

\subsection{HDW 5}

\begin{figure}
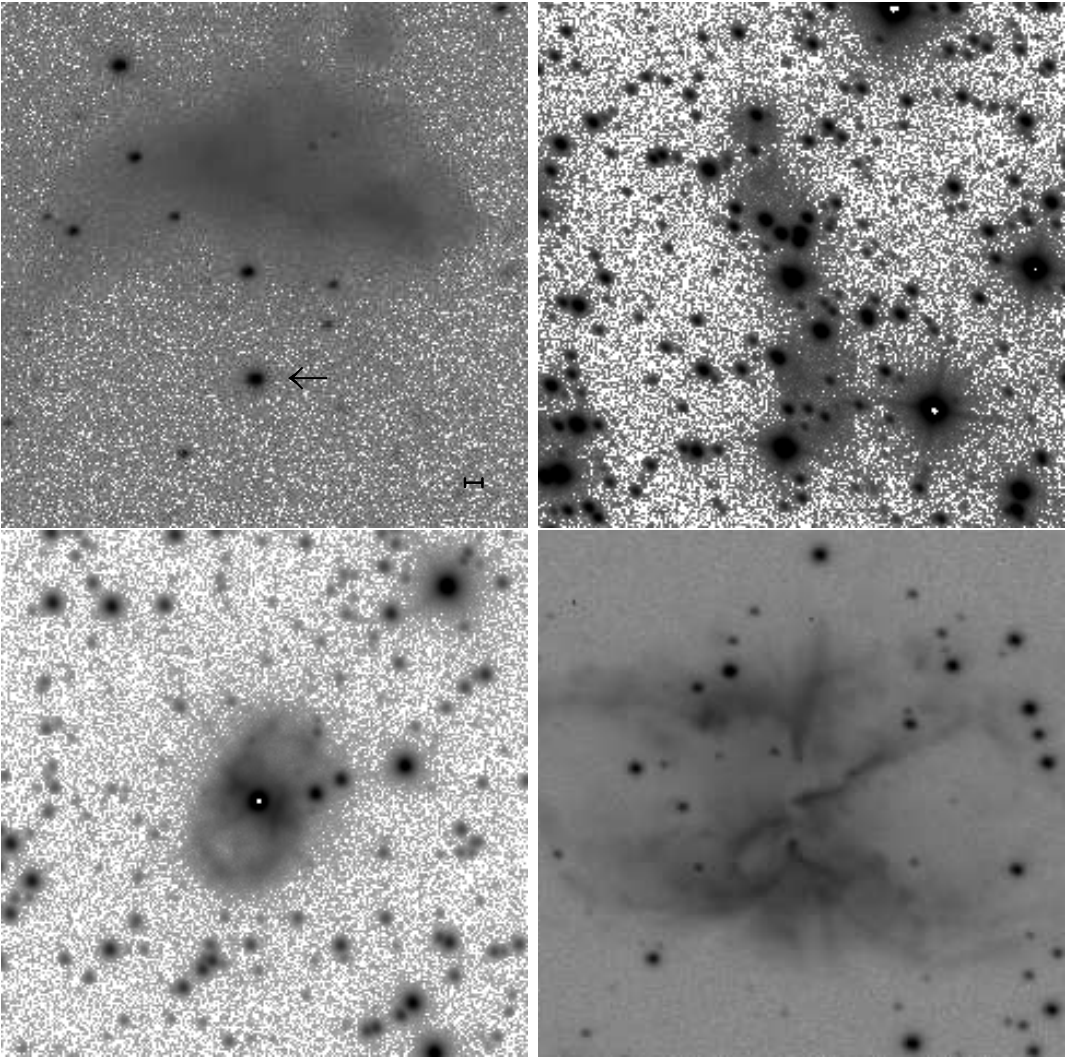

\includegraphics[angle=180, width=70mm]{f1a.eps}
\includegraphics[angle=180, width=70mm]{f1b.eps} \\
\includegraphics[angle=180, width=70mm]{f1c.eps}
\includegraphics[angle=180, width=70mm]{f1d.eps}
\caption{R-band image of HDW~5 (top left), He~2-25 (top right),
   He~2-36 (bottom left), and NGC~2818 (bottom right).  The images are
   $80''\times80''$ with north up and east to the left.  The central
   star of HDW~5 is indicated with an arrow.  The mark in the lower
   right corner of HDW~5 indicates the size of 8 pixels.}
\label{fg:imager1}
\end{figure}

\begin{figure}
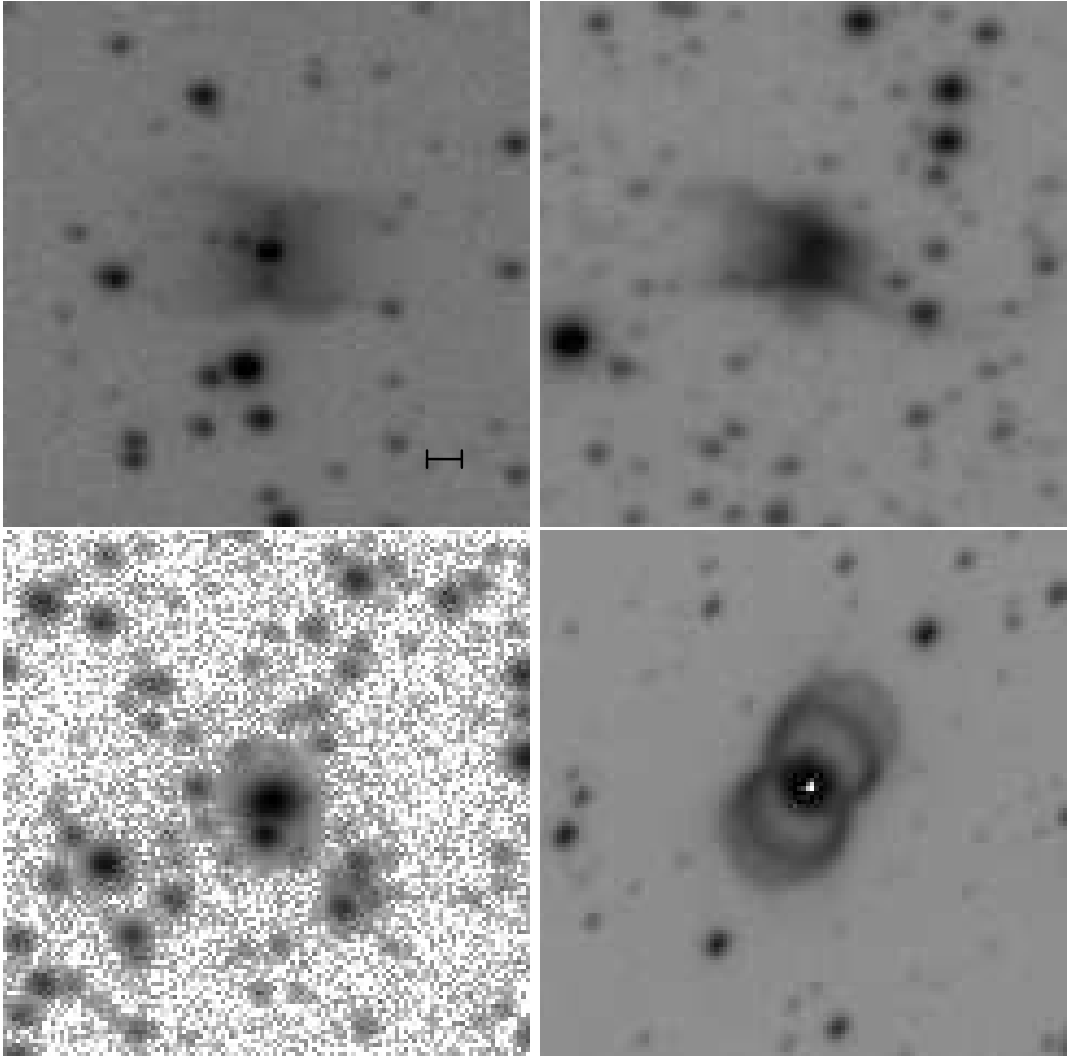

\includegraphics[angle=180, width=70mm]{f2a.eps}
\includegraphics[angle=180, width=70mm]{f2b.eps} \\
\includegraphics[angle=180, width=70mm]{f2c.eps}
\includegraphics[angle=180, width=70mm]{f2d.eps}
\caption{R-band image of He~2-64 (top left), He~2-123 (top right),
   He~2-186 (bottom left), and MyCn~18 (bottom right).
   The images are $40''\times40''$ with north up and east to the
   left.  The mark in the lower right corner of He~2-64 indicates the
   size of 8 pixels.}
\label{fg:imager2}
\end{figure}

\begin{figure}
\includegraphics[width=100mm]{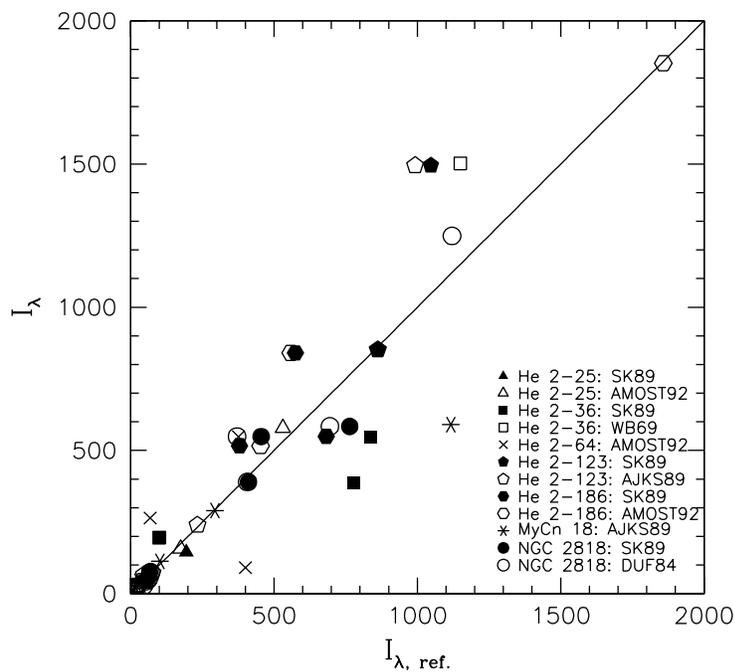} \\
\includegraphics[width=100mm]{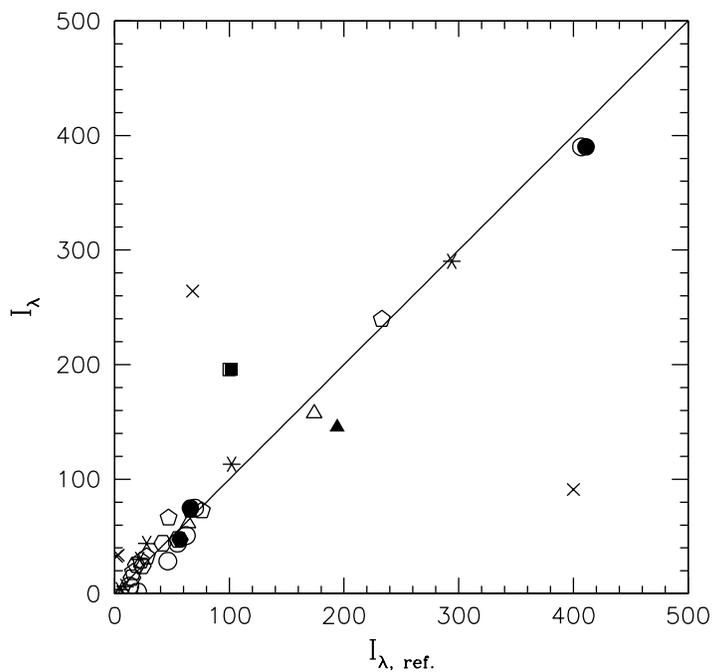}
\caption{Comparison of our nebular line intensities (I$_{\lambda}$,
  where I$_{4861}$=100) to others from the literature.  The solid line
  shows the 1:1 relation.  The bottom plot is a zoomed version of the
  top plot.  The references in the label of the top plot are: SK89:
  \cite{shaw89}, AMOST92: \cite{acker92}, WB69: \cite{webster69},
  AJKS89: \cite{acker89}, and DUF84: \cite{dufour84}. }
\label{fg:line_ratios}
\end{figure}

\begin{figure}
\includegraphics[angle=270, width=140mm]{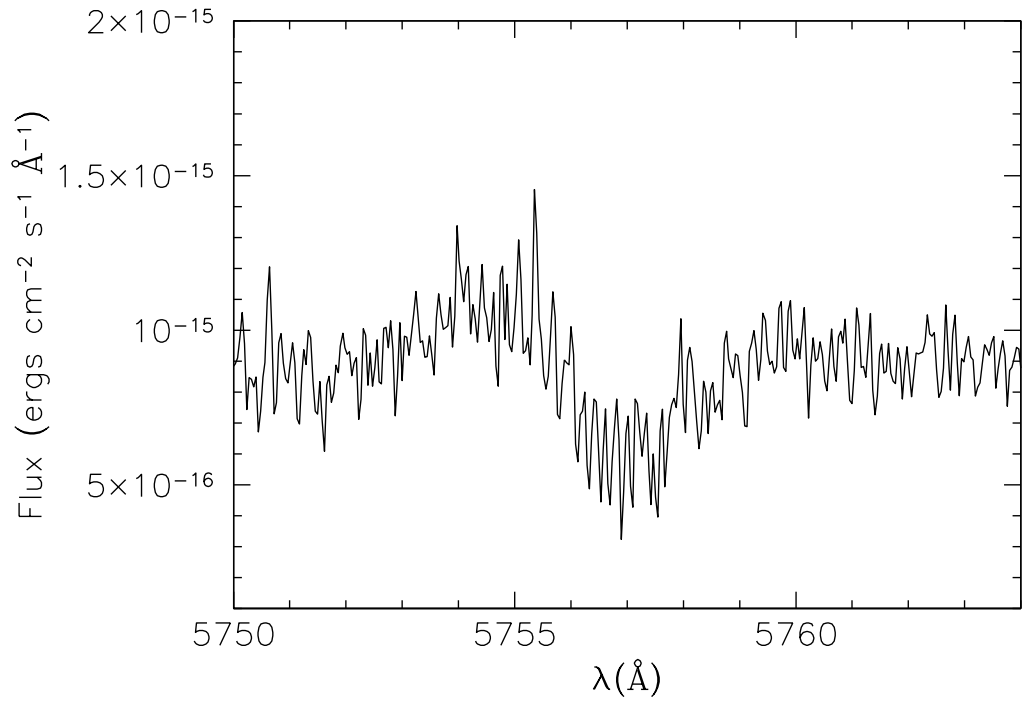}
\caption{An absorption feature in HDW~5.  The measured central
wavelength is at 5758~\AA.}
\label{fg:hdw5-stellar}
\end{figure}

  HDW 5 has a unique sickle-shape nebula (Figure \ref{fg:imager1}, top
  left).  It was discovered by \cite{hartl87} during a search for new
  PNe on Palomar Observatory Sky Survey (POSS).  Assuming a sphere
  with only the nebula visible in the north, \citeauthor{hartl87}
  proposed the blue star in the middle as the central star candidate
  (indicated with an arrow in Figure \ref{fg:imager1}).  \cite{ali99}
  studied this nebula using its H$\alpha$ and [\ion{N}{2}] narrow band
  images and low-dispersion optical spectra and concluded that this
  nebula is in very low excited class and its low He abundance is
  consistent with type II PNe as defined by \cite{peimbert83}.
  Because of its peculiar shape, there is also some speculation that
  HDW~5 is not a PN, pending further analysis (Frew 2006, private
  communication).  \cite{mendez91} tentatively classified the central
  star candidate as hgO(H) -- a high gravity star with very broad
  Balmer absorptions.  The spectrum that we obtained shows no hydrogen
  lines.  Only one unidentified absorption line is found at 5758\AA~as
  shown in Figure \ref{fg:hdw5-stellar}.  Thus we are not able to
  assign a spectral class for this CS candidate.

\subsection{He 2-25}

  The nebula has a pair of bipolar lobes with a stellar core (Figure
  \ref{fg:imager1}, top right).  It was discovered by \cite{henize67}
  during an H$\alpha$ survey.  \cite{corradi95} noted that the
  spectral characteristics of this nebula are common to a subclass of
  bipolar PNe, which includes the well-studied M~2-9 and has
  properties such as: (i) highly collimated nebulae; (ii) unusually
  high core densities, (iii) H$\alpha$ profiles with extended wings
  and self-absorption features; and (iv) rich \ion{Fe}{2} and
  [\ion{Fe}{2}] emission in generally low-ionization spectra.  Those
  characteristics closely resemble those of interacting binary systems
  such as symbiotic stars.  However, \cite{corradi95} found no
  direct evidence of a binary system as the central source of He~2-25.
  We did not detect stellar emission or absorption lines for this
  object.

\subsection{He 2-36}

  The image of He 2-36 displays a bright central source and an
  equatorial band (Figure \ref{fg:imager1}, bottom left).  A pair of
  arcs extend from the edges of the equatorial band toward the north
  and south, making it an S-shape nebula.  The observed spectrum shows
  strong hydrogen absorption lines and many metal lines (e.g., iron
  and titanium), characteristic of an A-type star.  The CS has been
  previously classified as type A2~III \citep[e.g.,][]{mendez78}.  It
  has been shown that the central star is a binary system consisting
  of a type A2~III star with a hot companion, the hot companion being
  the source of ionizing flux, as detected in the UV spectrum acquired
  with the IUE \citep{feibelman01}.  The optical spectrum of He~2-36
  is discussed in more detail in Section \ref{sec:split}.

\subsection{NGC 2818}

  This nebula is a member of an open cluster with the same name
  \citep{tifft72}.  This provides an accurate distance of the PN
  \citep[][and references therein]{phillips98}.  \cite{phillips98}
  presented narrow band images of this nebula and confirmed the
  complexity of its bipolar outflow structure (Figure
  \ref{fg:imager1}, bottom right) by previous authors
  \citep[e.g.,][]{dufour84,banerjee90}.  No stellar emission or
  absorption lines were detected from the central star.  The stellar
  continuum is very faint.  It may be an indication that the central
  star is not at the center of the nebula, where the observation was
  pointed.

\subsection{He 2-64}

\begin{figure}
\includegraphics[angle=270, width=80mm]{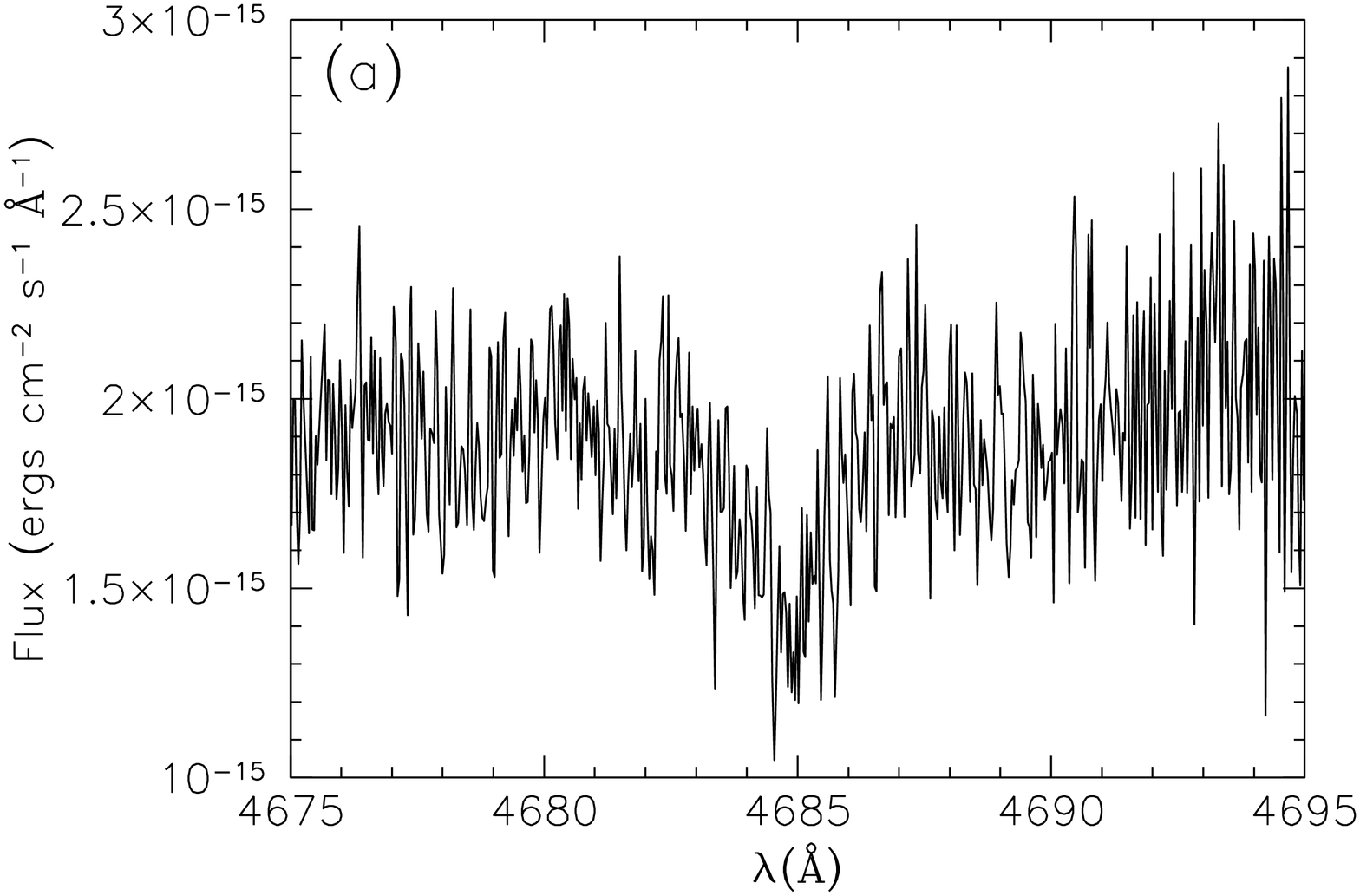}
\includegraphics[angle=270, width=80mm]{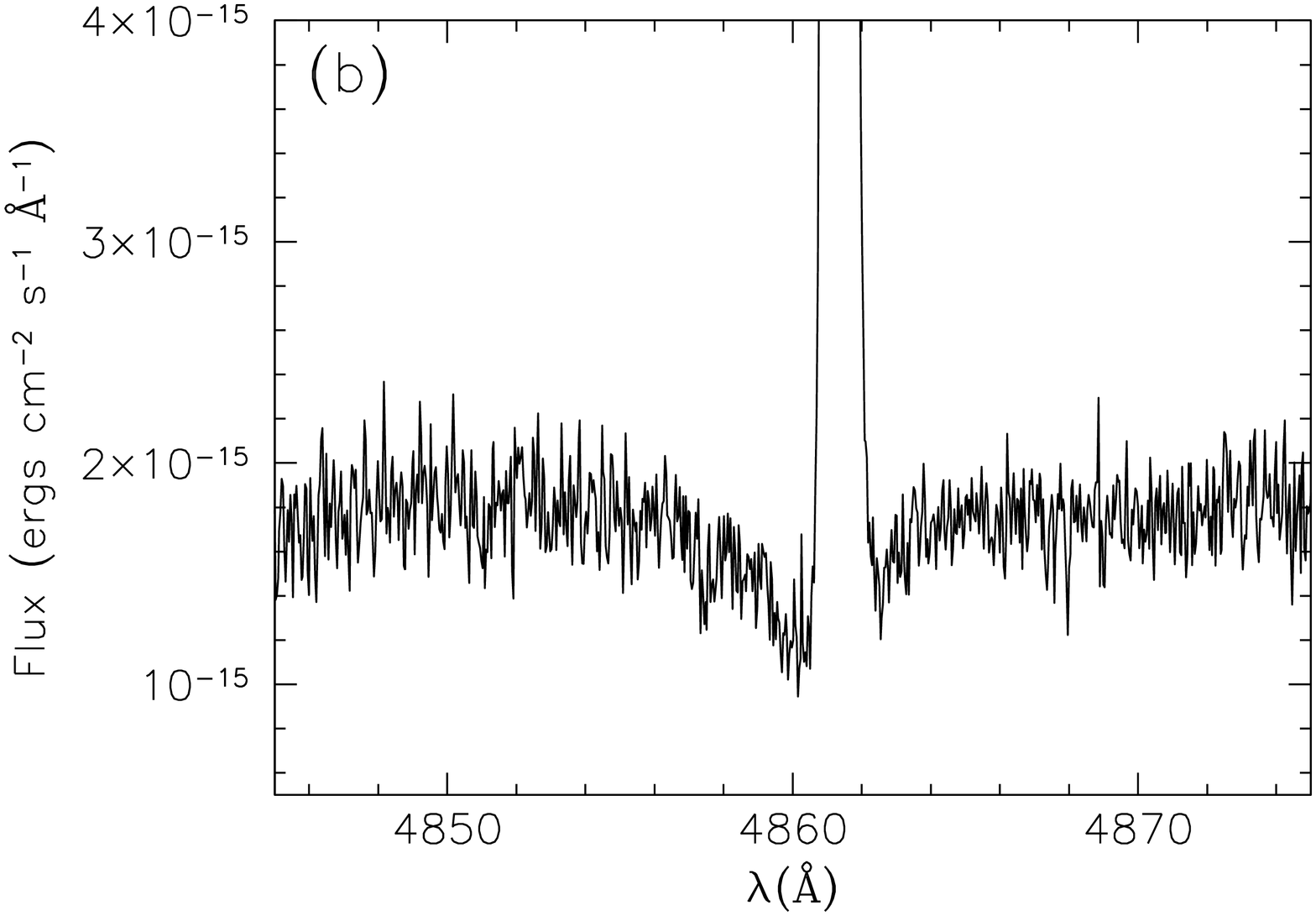}
\includegraphics[angle=270, width=80mm]{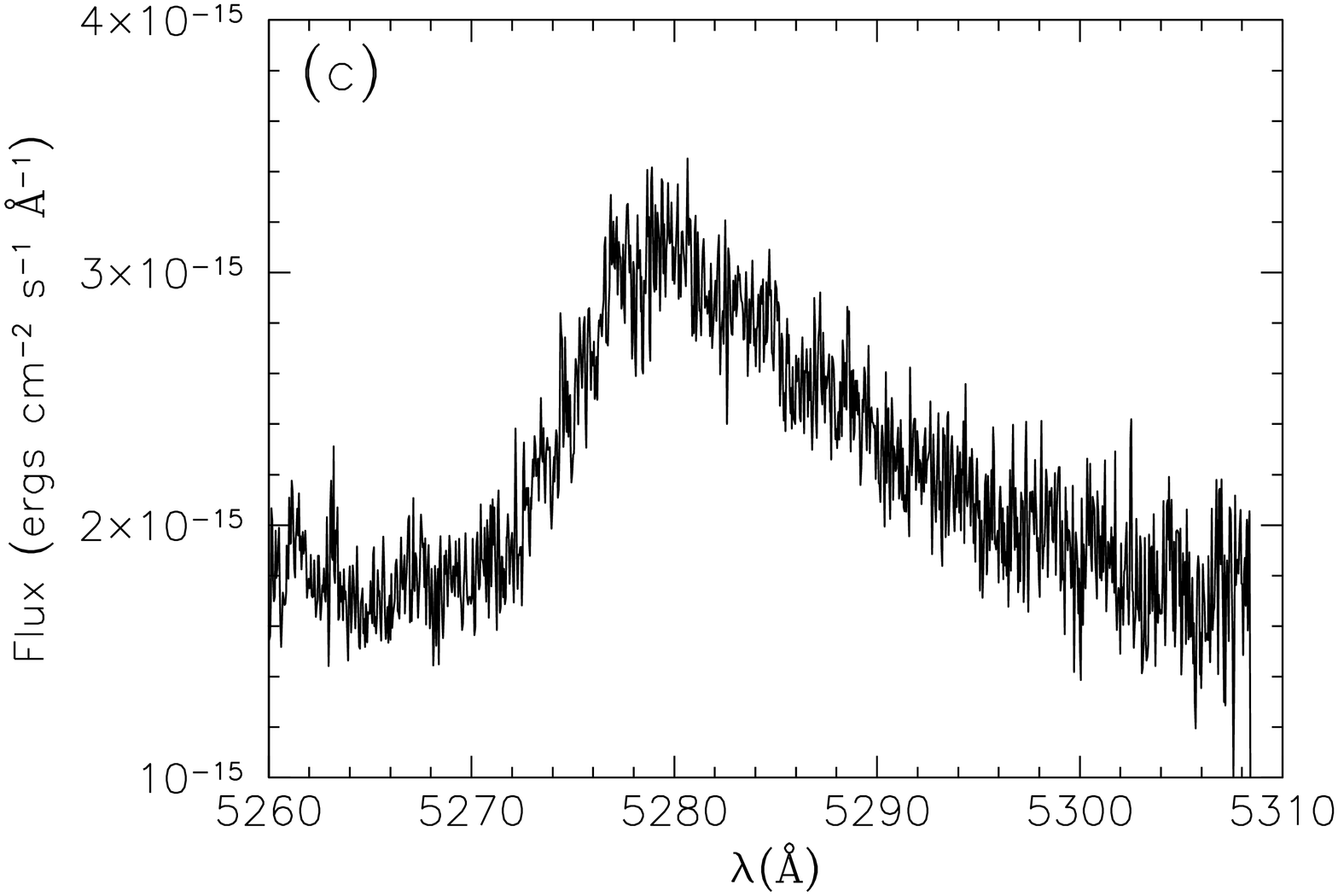}
\includegraphics[angle=270, width=80mm]{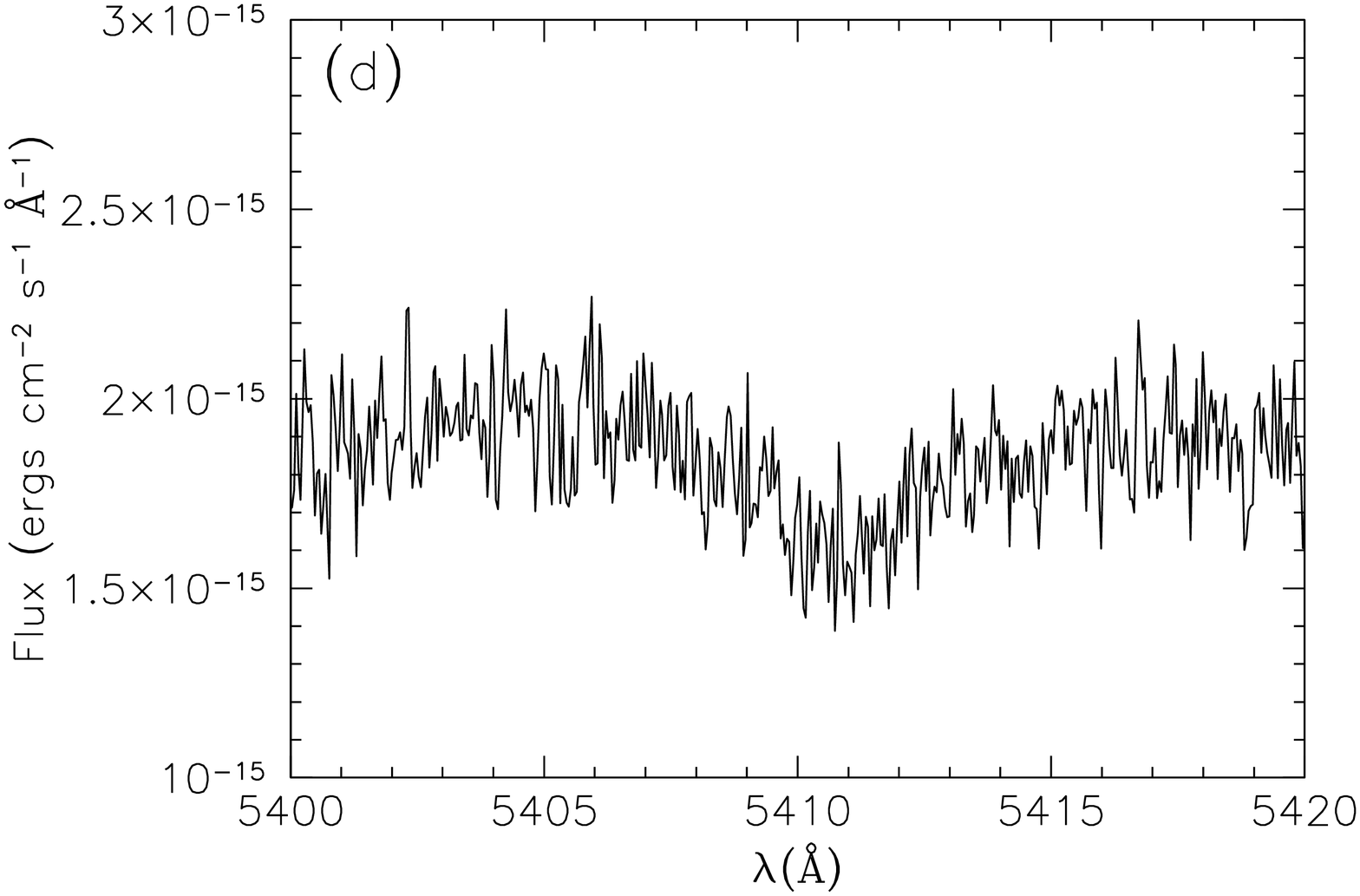}
\includegraphics[angle=270, width=80mm]{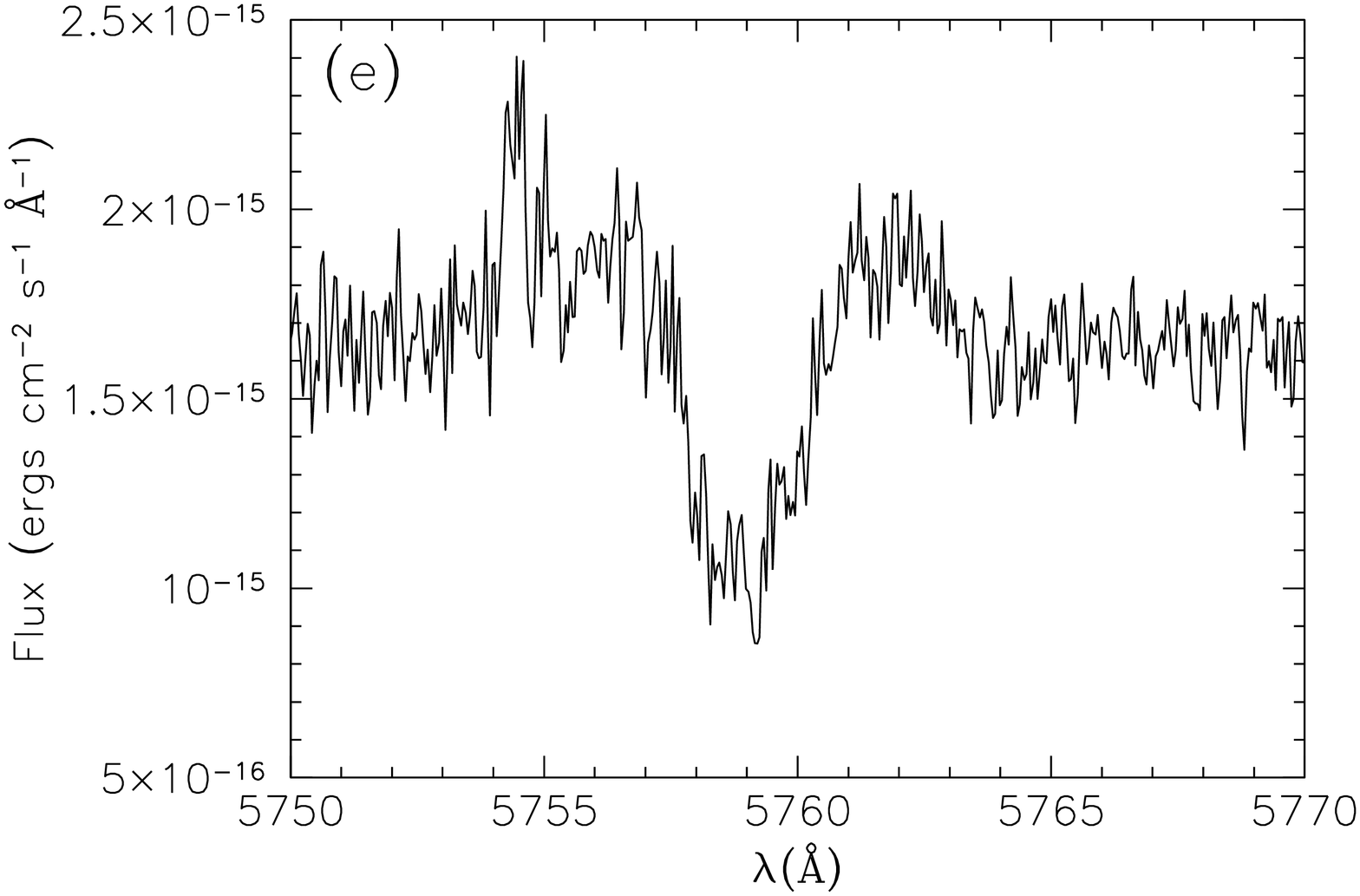}
\caption{Spectrum of He~2-64: (a) \ion{He}{2} absorption at 4686~\AA;
  (b) H$\beta$ $\lambda$4861 stellar absorption and nebular emission
  line; (c) A very broad emission feature at 5281~\AA~in He~2-64; (d)
  \ion{He}{2} absorption at 5411~\AA; (e) An absorption feature at
  5759~\AA~in He 2-64.}
\label{fg:he2-64.line}
\end{figure}

  The image of He 2-64 (Figure \ref{fg:imager2}, top left) shows a
  bright central source with a pair of lobes.  It is classified as a
  bipolar PN \citep[e.g.,][]{corradischwarz95}.  The spectrum of the
  central star shows a few stellar absorption lines, including
  \ion{He}{2} $\lambda$4686~and \ion{He}{2} $\lambda$5411 (Figure
  \ref{fg:he2-64.line}).  We classify this star as O(He), according to
  the scheme outlined by \cite{mendez91}.

\begin{figure}
\epsscale{.8}
\plotone{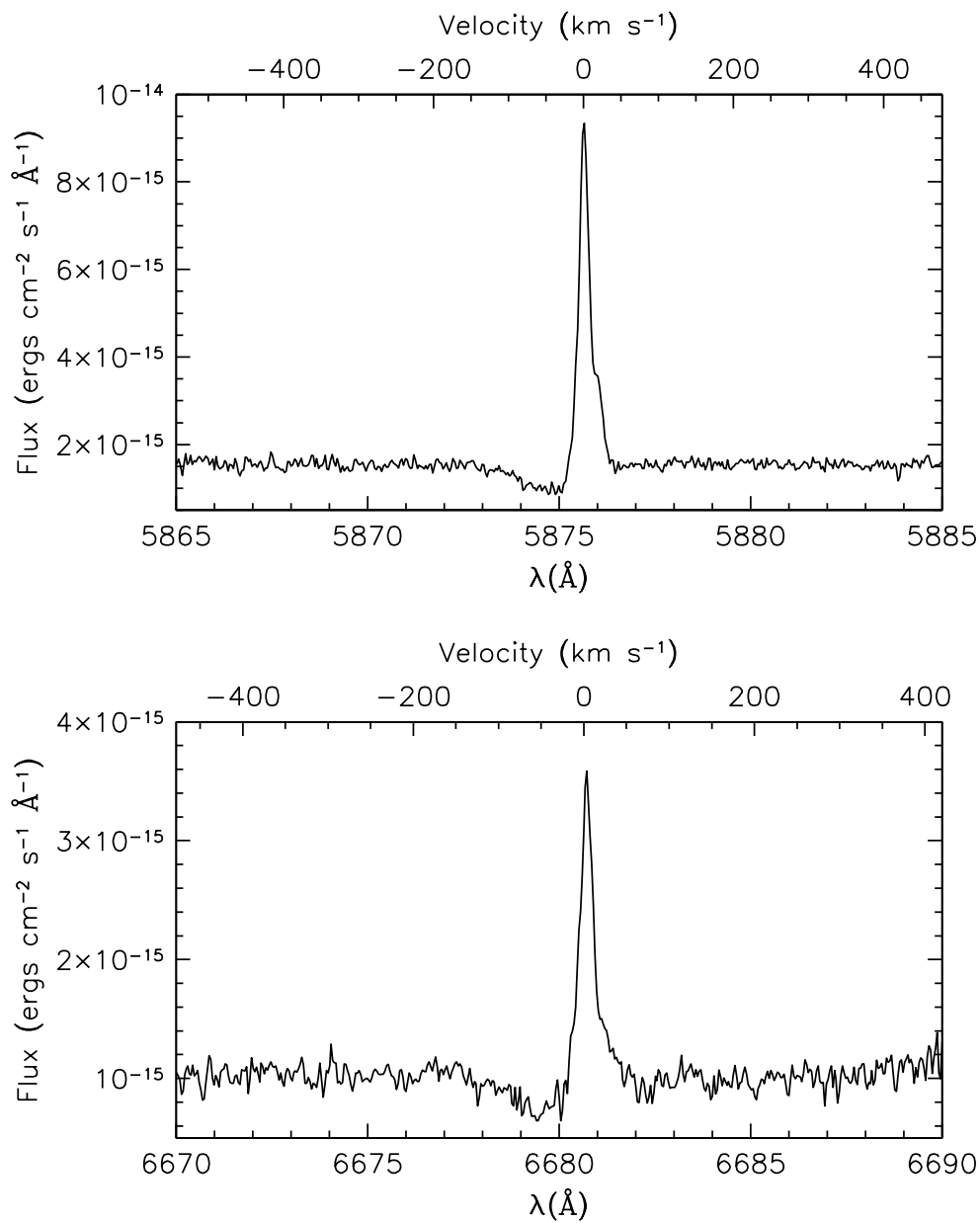}
\epsscale{1}
\caption{Spectrum of He~2-64: P-Cygni profiles.  The upper panel shows
  He I $\lambda$5876 \AA, the lower panel shows He I $\lambda$6678
  \AA.  The expansion velocity inferred from these profiles is in the
  order of $\sim$ 100 km s$^{-1}$.  The velocity scale at the top of
  each panel is set to 0 at the peak of the emission.}
\label{fg:he2-64.p-cygni}
\end{figure}

  The spectrum of He~2-64 shows P-Cygni profiles of the \ion{He}{1}
  lines, most clearly shown at $\lambda$5876 and $\lambda$6678 (Figure
  \ref{fg:he2-64.p-cygni}).  The properties of these lines are listed
  in Table \ref{tb:p-cygni}.  The velocity deduced from the P-Cygni
  profiles is of the order $\sim$ 100 km s$^{-1}$.  This finding
  indicates that the star has an extensive expanding atmosphere.  The
  P-cygni profiles are more clearly seen in the 4-pixel extraction,
  suggesting that they are partly masked by the nebular emission.  In
  the optical regime, P-Cygni profiles are more often seen in the
  \ion{He}{2} lines \citep[e.g.,][]{mendez88,mendez90}.  For example,
  \cite{phillips05} reported P-Cygni profiles in the \ion{He}{2}
  $\lambda$4859 line of A78, with a terminal velocity $\sim 4 \times
  10^3$ km s$^{-1}$.

\subsection{He 2-123}

\begin{figure}
\includegraphics[angle=270, width=80mm]{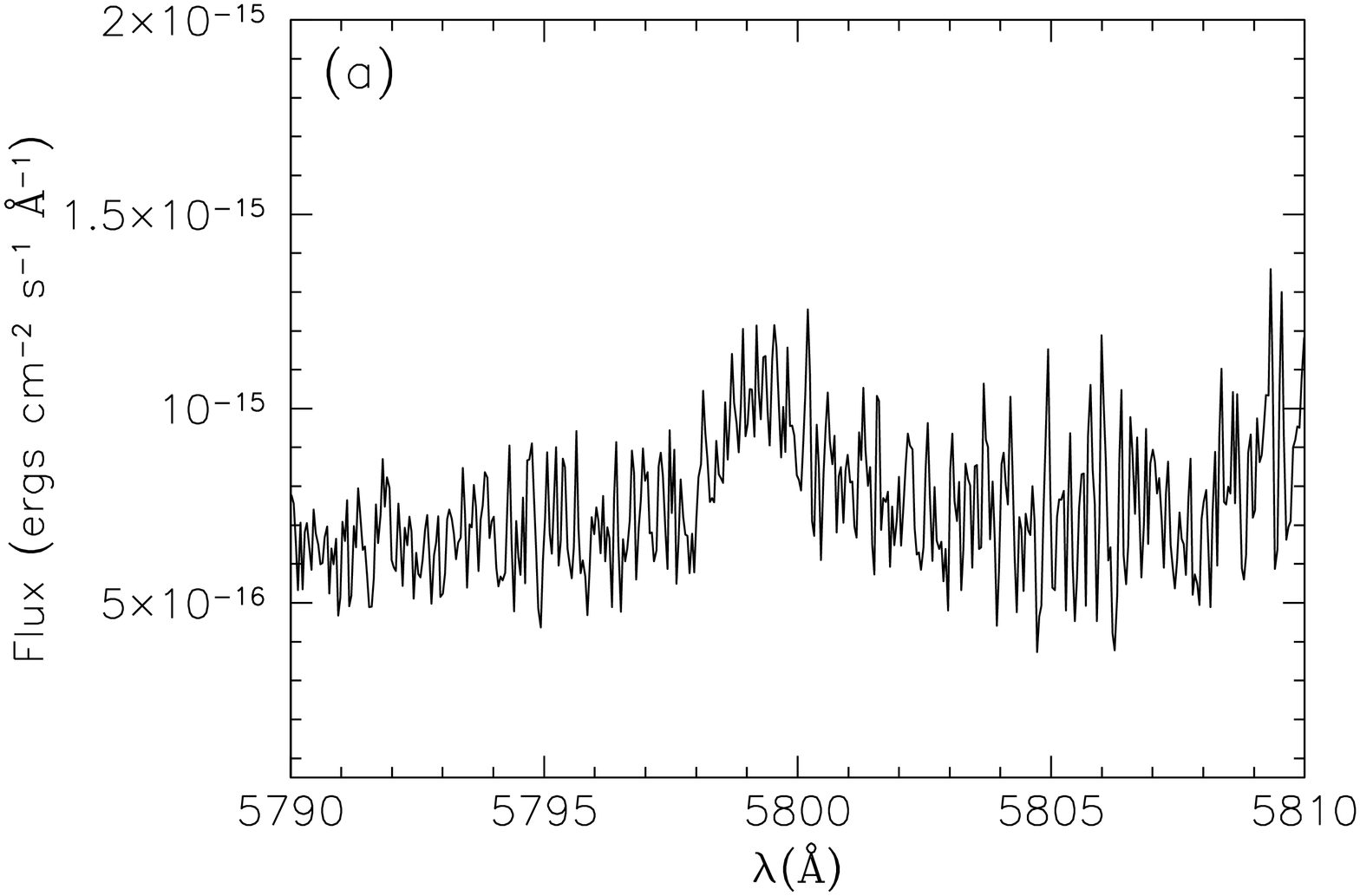}
\includegraphics[angle=270, width=80mm]{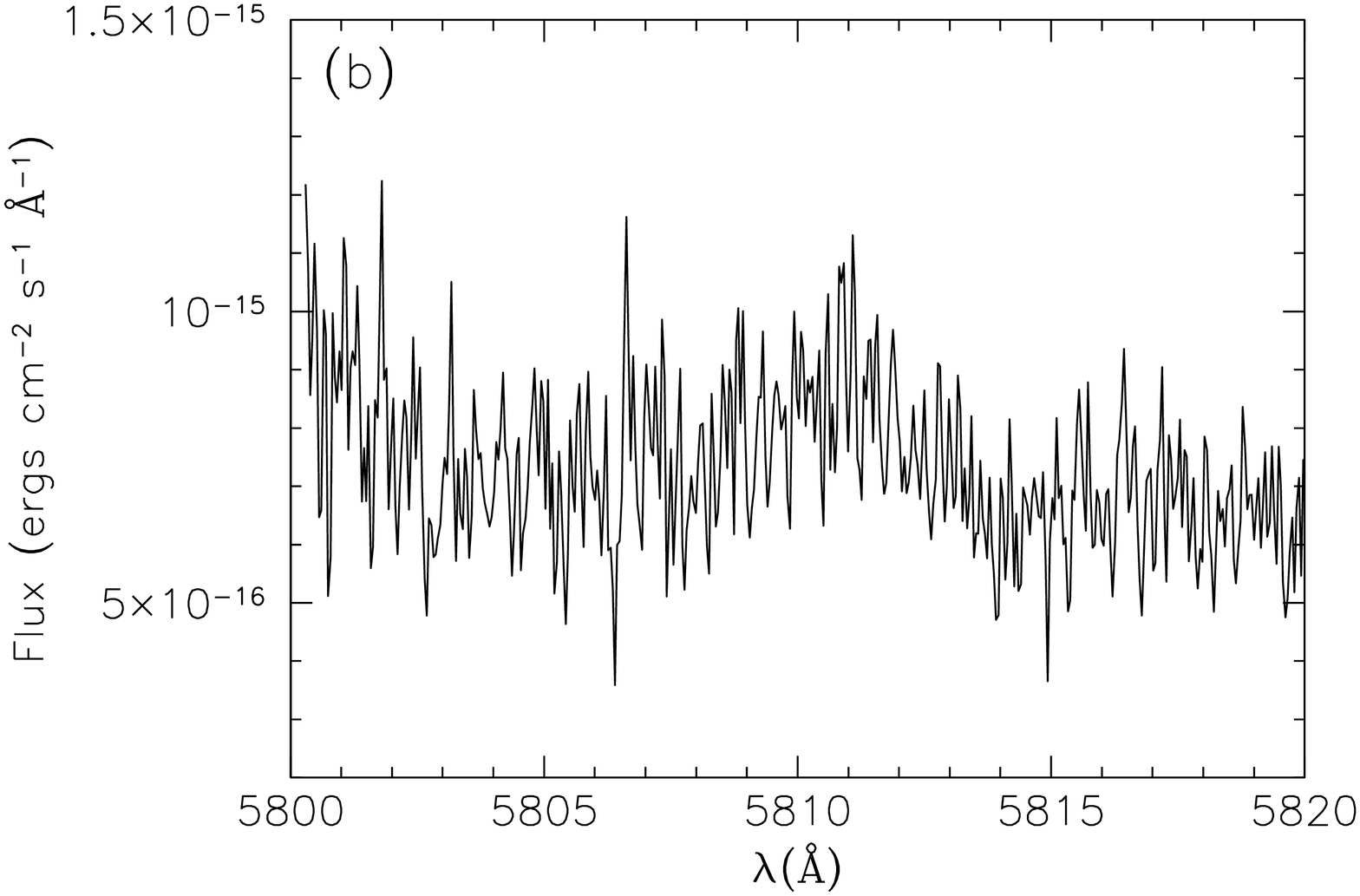}
\caption{Spectrum of He~2-123.  Stellar emission line of \ion{C}{4} at
  (a) 5801~\AA~and (b) 5811~\AA.}
\label{fg:he2-123-stellar}
\end{figure}

  He 2-123 (Figure \ref{fg:imager2}, top right) shows an equatorial
  waist with a pair of bipolar lobes extending to the east and the
  west.  This PN was discovered in \citeauthor{henize67}'s
  \citeyearpar{henize67} H$\alpha$ survey.  \cite{corradi93a} first
  noticed that its lobes have point symmetrical brightness
  distribution.  We have detected two weak emission lines from the
  central star: \ion{C}{4} $\lambda$5801 and $\lambda$5811 (Figure
  \ref{fg:he2-123-stellar}).  At the position of \ion{He}{2}
  $\lambda$4686, there seems to be some stellar emission, but it is
  too faint to be measured.  If this \ion{He}{2} emission is real, the
  spectral type of the central star would be Of(C) or Of-WR(C).  The
  faint \ion{C}{4} emission lines also fit into the category of the
  weak emission-line (WEL) CSPN as defined by \cite{tylenda93}.

\subsection{He 2-186}

  The nebula (Figure \ref{fg:imager2}, bottom left) has several bright
  arcs of emission in the inner region, with a pair of low-ionization
  knots about 5$''$ from the core \citep[see Figure 4
  of][]{corradi00}.  He 2-186 is a small, poorly studied PN.  The
  H$\alpha$ emission was first studied by \cite{henize67}.  The narrow
  band images presented by \cite{schwarz92} first revealed the
  existence of the point-symmetric knots.  \cite{corradi00} performed
  a detailed morphological and kinematic study of this nebula and
  discussed the possibility that these high-velocity knots were the
  result of precessing outflows from the central star.
  \cite{corradi00} also speculated the existence of binary system as
  the central star.  However, no direct evidence of it was found.  The
  spectrum of the central star shows no stellar emission or absorption
  lines.

\subsection{MyCn 18}

  MyCn 18, the 'Engraved Hourglass' nebula (Figure \ref{fg:imager2},
  bottom right), was originally observed by \cite{mayall40}.  This PN
  shows an extreme bipolar morphology, with a narrow pinched waist and
  an open-ended hourglass structure.  Because of its striking shape,
  the nebula has been studied in a wide range of wavelength and the
  nature of its central star has been discussed by various authors
  \citep[][and references therein]{bryce04}.  \cite{corradi93b}
  compared this object to the symbiotic bipolar nebulae He 2-104 and
  BI Crucis, based on their morphological similarity, but noted no
  direct evidence that MyCn~18 has a central symbiotic system.
  \cite{bryce97} and \cite{oconnor00} discussed several possible
  mechanisms that can produce the bipolar, knotty outflow seen in
  MyCn~18 and favored a nova-like ejection from a central binary
  system.

\begin{figure}
\includegraphics[angle=270, width=80mm]{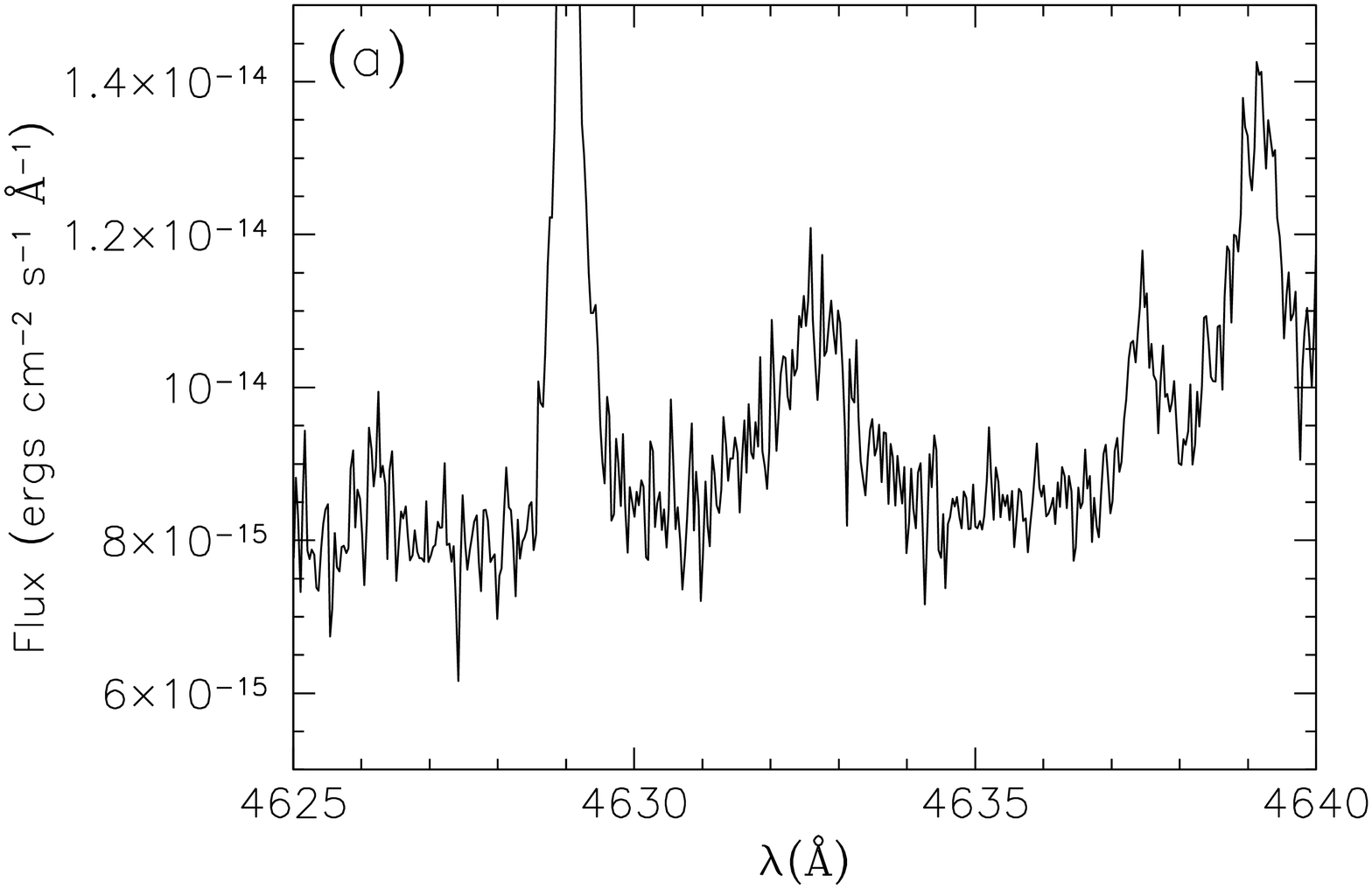}
\includegraphics[angle=270, width=80mm]{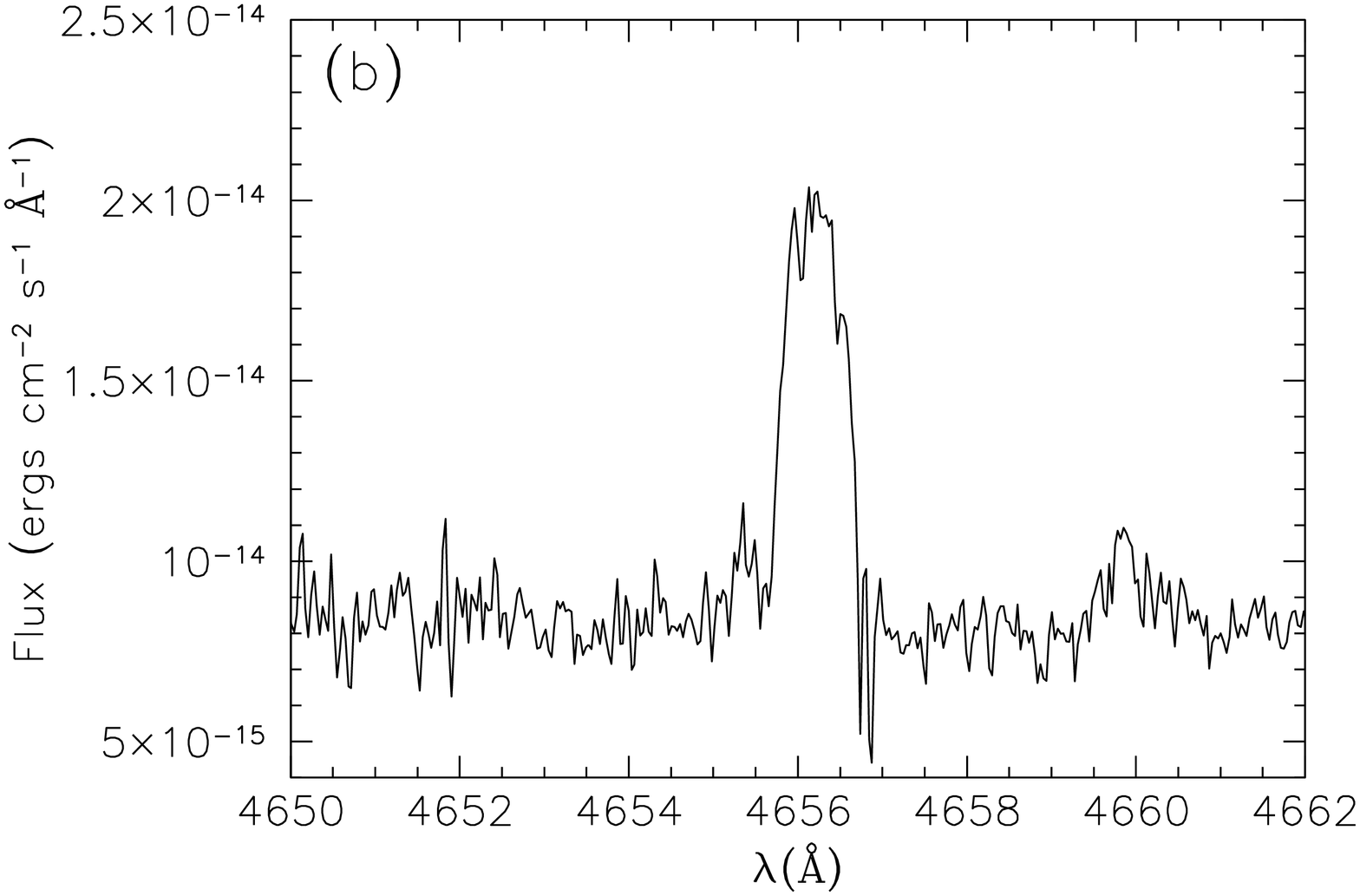}
\includegraphics[angle=270, width=80mm]{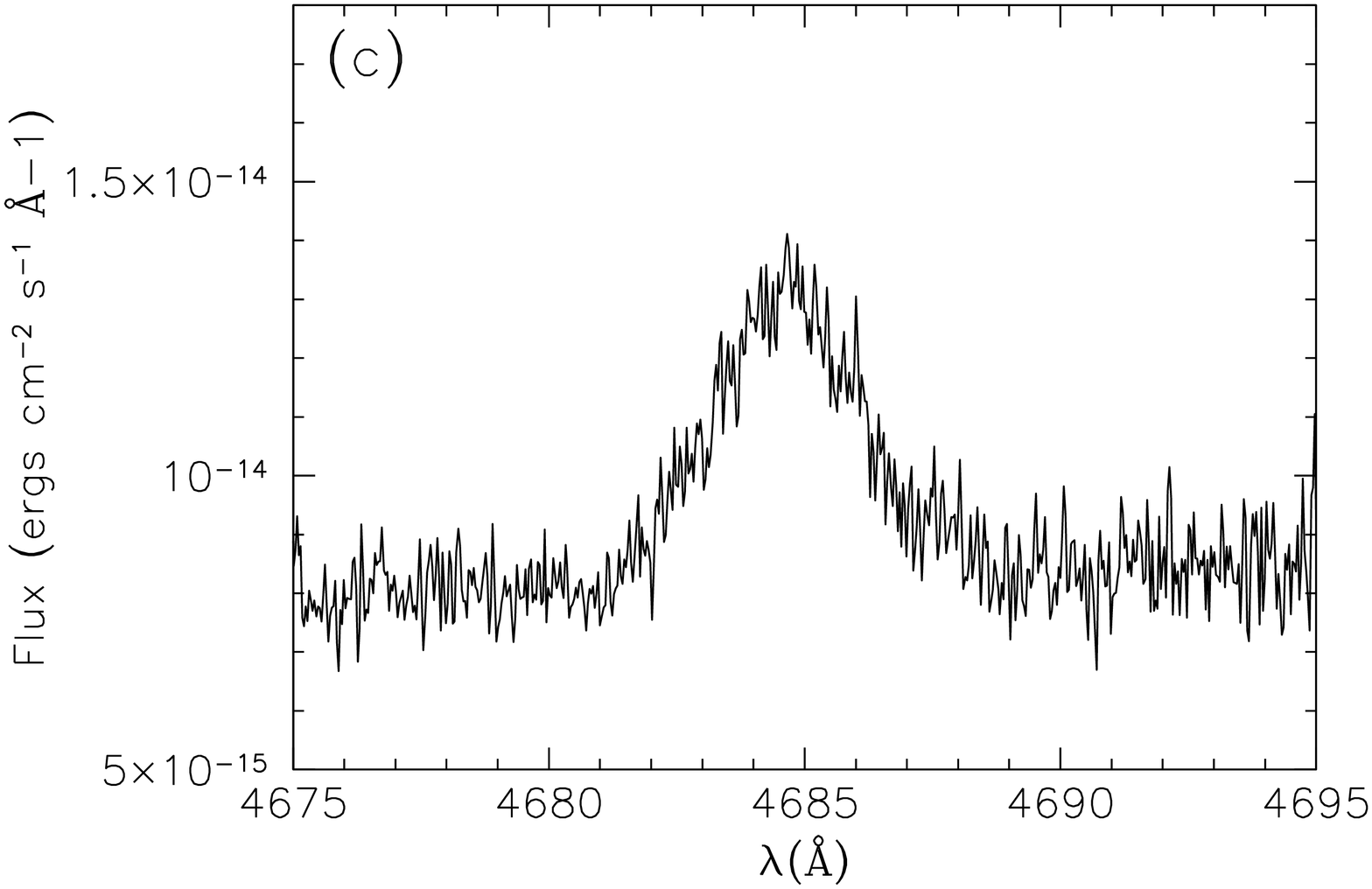}
\includegraphics[angle=270, width=80mm]{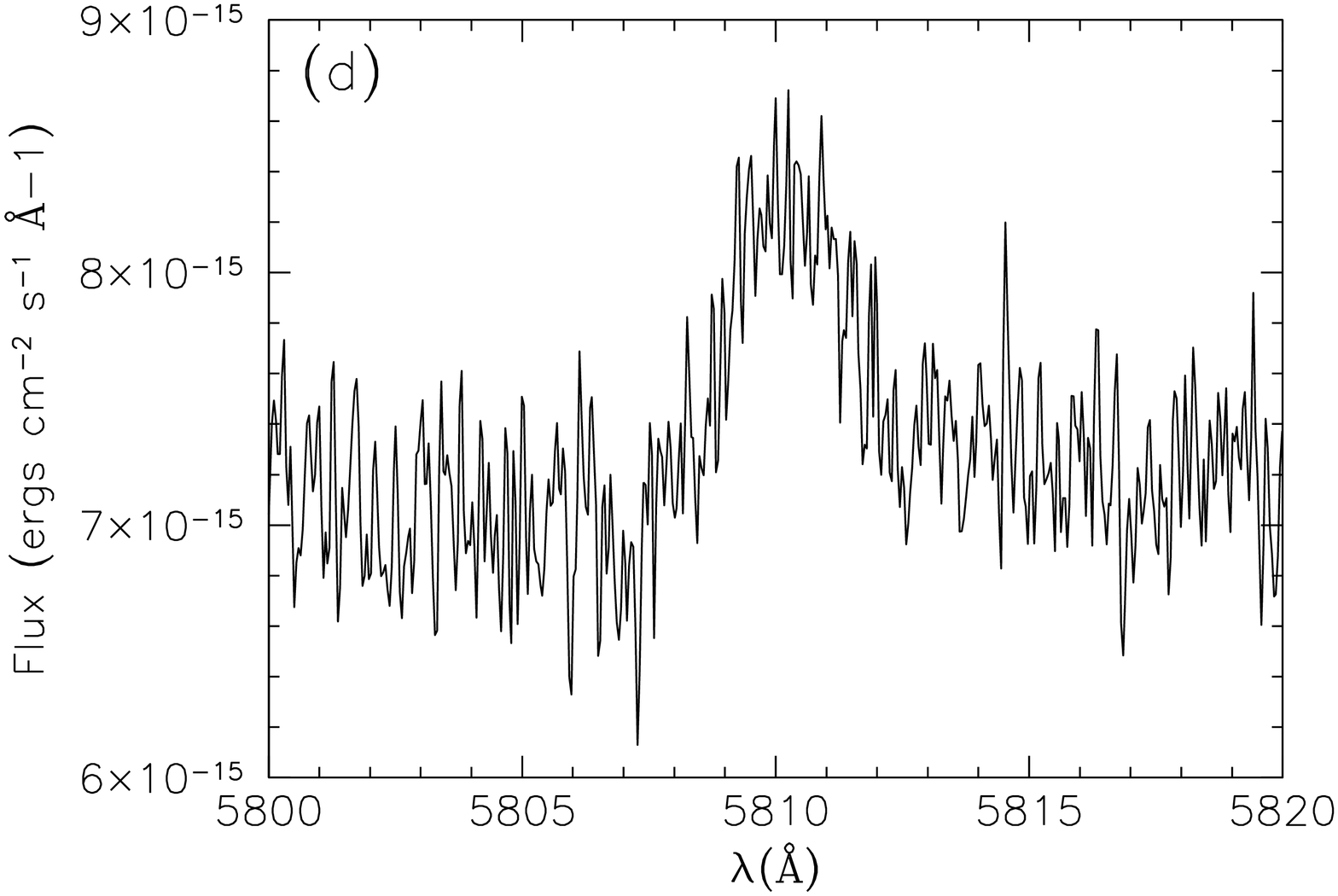}
\includegraphics[angle=270, width=80mm]{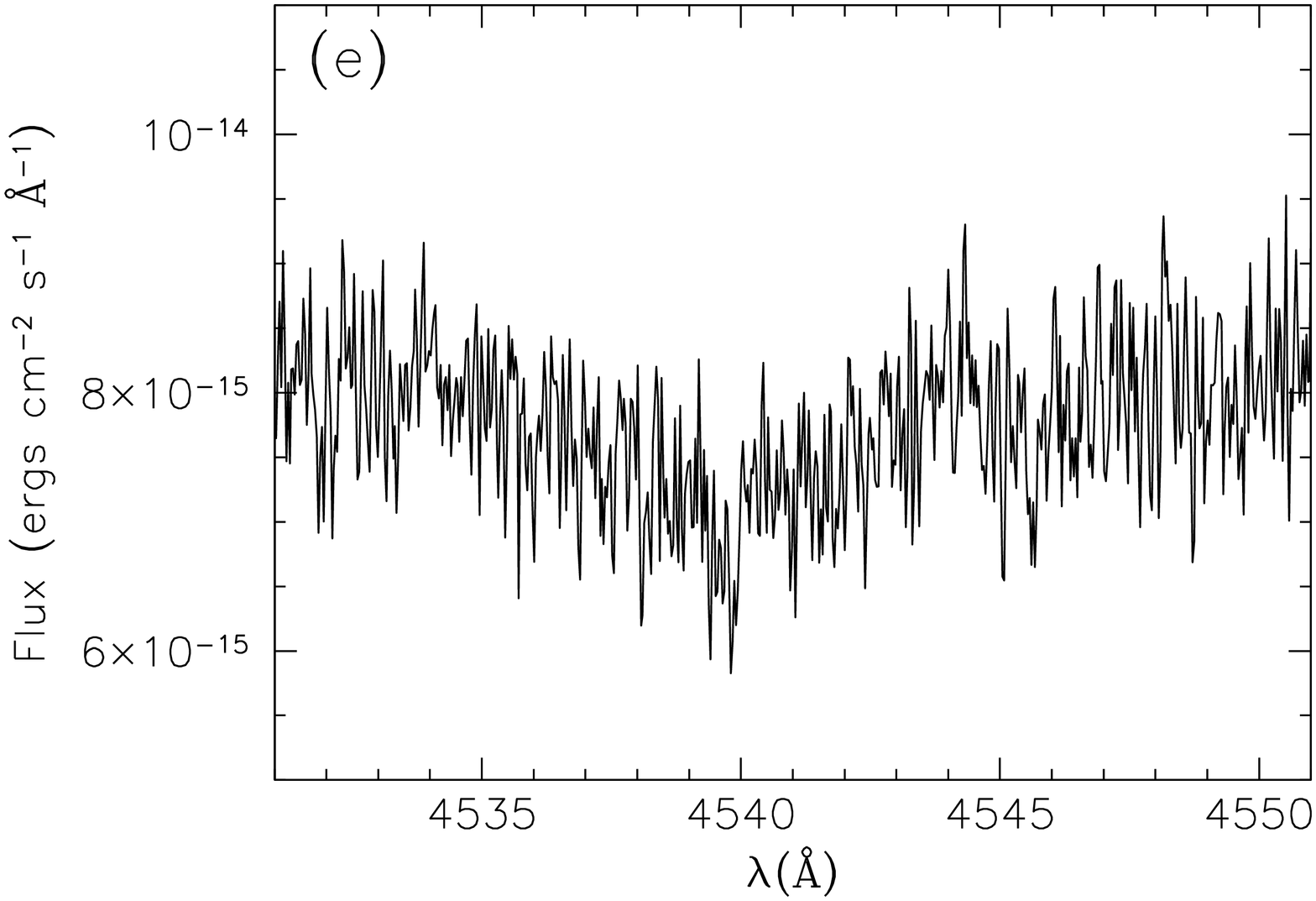}
\includegraphics[angle=270, width=80mm]{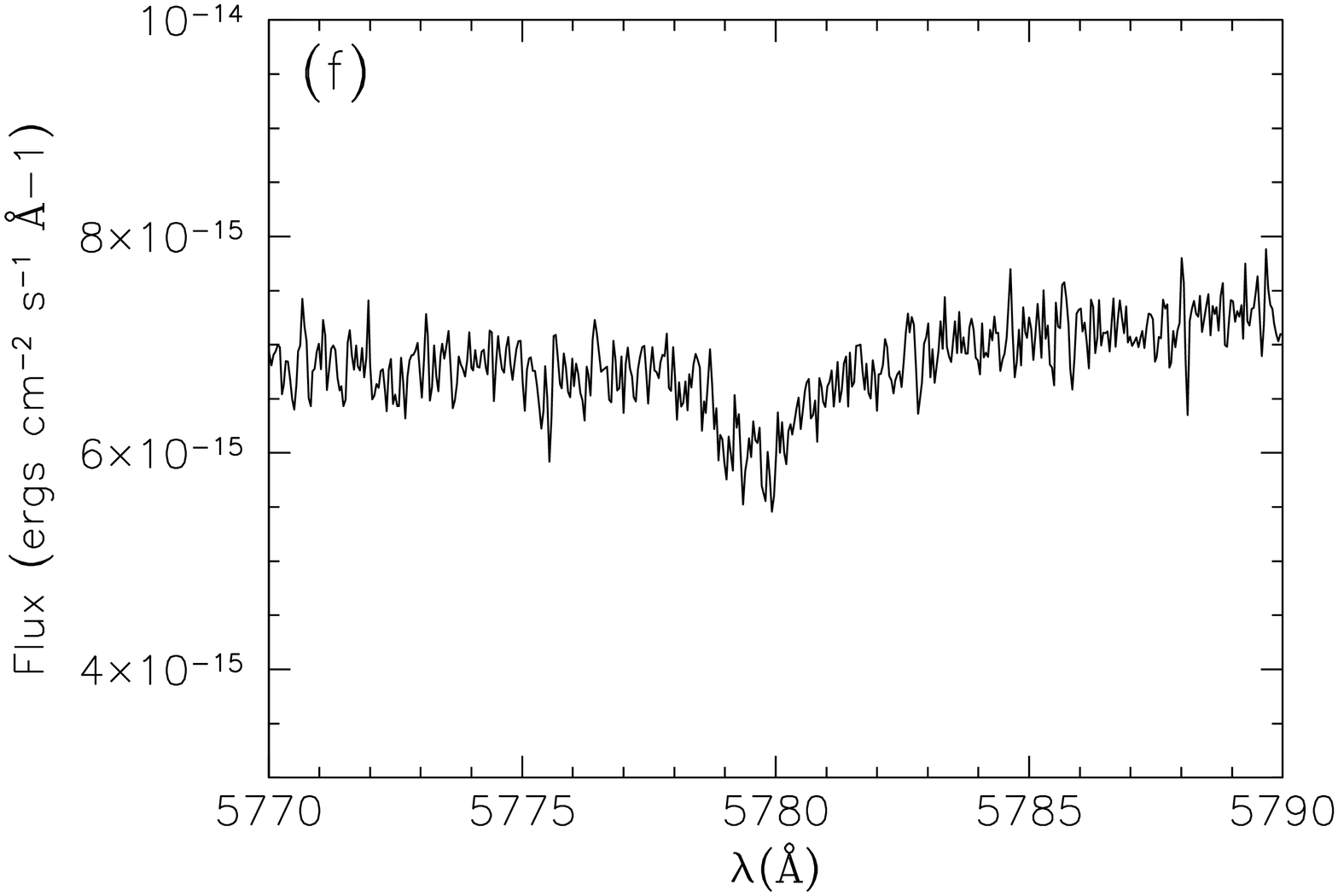}
\caption{Spectrum of MyCn~18: (a) \ion{N}{3} emission at 4634~\AA; (b)
\ion{C}{4} emission at 4658~\AA; (c) \ion{He}{2} emission at 4686~\AA;
(d) \ion{C}{4} emission at 5811~\AA; (e) \ion{He}{2} absorption at
4541~\AA; (f) A faint absorption line at 5780~\AA.}
\label{fg:mycn18.line}
\end{figure}

  The spectrum of the central star shows four emission lines:
  \ion{N}{3} $\lambda$4634, \ion{C}{4} $\lambda$4658, \ion{He}{2}
  $\lambda$4686, and \ion{C}{4} $\lambda$5811.  It also shows one very
  faint absorption line (\ion{He}{2} $\lambda$4541), and one
  absorption line, unidentified, at $\lambda$5780.  All these features
  are shown in Figure \ref{fg:mycn18.line}.  \cite{mendez91}
  classified its spectral type as Of(H) -- \ion{He}{2} $\lambda$4686
  is a narrow emission (FWHM$<4$\AA) with H$\gamma$ in absorption.  We
  found that the \ion{He}{2} $\lambda$4686 line has a FWHM of 3.7\AA.
  However, we did not find any Balmer absorption lines for the central
  star.  Therefore, we classify the spectra as Of(C).

\section{Search for Zeeman-Split Stellar Lines}
\label{sec:split}

  If a Zeeman split of magnetic origin is detected in the spectrum,
  the field strength could be inferred with:
\begin{equation}
\label{eq:B}
  \Delta \lambda = 4.67 \times 10^{-13} g_{\rm eff} \lambda_0^2 B
\end{equation}
  where $\Delta \lambda$ is the split separation, $\lambda_0$ is the
  wavelength of the spectral line, $g_{\rm eff}$ is the effective
  Land{\'e} factor and B is the magnetic field strength in Gauss
  \citep{leone03}. Equation (\ref{eq:B}) can be also used to estimate
  upper limits to the magnetic field.

  Among the 8 CSs observed, only the spectra of He~2-36, He~2-64, and
  MyCn~18 have high enough S/N to determine whether their stellar
  lines might be Zeeman split.  We closely examined the stellar
  spectra of these three objects to look for signs of splitting due to
  the presence of a magnetic field.  A possible split was identified
  visually if a stellar line shows more than one peak in emission or
  more than one dip in absorption.  If the feature shows two apparent
  components, and (e.g. for emission) the intensity drop between the
  components is greater than 5 times the 1-sigma noise level of the
  spectrum, we concluded that the feature is real and performed
  further analysis on the feature.

  No evidence of Zeeman splitting was found in the spectra of He~2-64
  and MyCn~18.  We could place an upper limit to the magnetic field
  associated with these two CSs using Equation (\ref{eq:B}).  With an
  optimal spectral resolution of 0.1~\AA, the minimum detection of the
  wavelength displacement is 0.05~\AA.  To cause a displacement of
  0.05~\AA~in a simple Zeeman triplet ($g_{\rm eff}=1.0$), the field
  strength has to be $\sim$ 5000 G.  However, due to the nature of our
  data -- faint central stars thus low S/N for stellar continuum
  ($\sim$ 10 to 20 for He~2-64 and MyCn~18), we believe that we are
  unable to detect a split feature with a separation of less than
  0.4~\AA.  This translates to a field strength of $\sim$ 20,000
  G.  Therefore no magnetic field stronger than 20,000 G was
  found in the central stars of He~2-64 and MyCn~18.

  The spectrum of He~2-36 is more difficult to interpret.  We know
  that this star has two components, the cool companion being an A2
  III star, and the hot companion detected to date only in the
  ultraviolet spectra.  The prominent features of our He~2-36 spectrum
  are those metal lines expected in a A2 III stellar spectrum, such as
  those of HD~210111 (an A2 III-IV star observed with the VLT, from
  the UVES Paranal Observatory Project, ESO DDT Program ID 266.D-5655,
  Bagnulo et al.\ 2003).  We have successfully fitted the Balmer lines
  in our spectrum by using the models by \citeauthor{kurucz92}
  \citeyearpar[][with the latest modifications from his
  website\footnote{http://kurucz.harvard.edu/}]{kurucz92} with $T_{\rm
  eff}$ = 8500 K and $\log g = 2$. There is no need to invoke a hot
  component to justify these lines.  On the other hand, when comparing
  He~2-36 with the spectrum of HD~210111, we were also able to
  identify two absorption lines at $\lambda$4686 and $\lambda$5876,
  where no metal lines typical of an A2 III star are expected.  We
  believe these two lines correspond to \ion{He}{2} $\lambda$4686 and
  \ion{He}{1} $\lambda$5876.  We show these two absorption lines in
  Figure \ref{fg:he2-36.4686} and \ref{fg:he2-36.5876}.  The profile
  of \ion{He}{2} is especially compelling.  The nebular emission line
  shows an asymmetric profile in the 8-pixel extraction, indicating
  the existence of the absorption.  In the 4-pixel extraction, the
  absorption is shown along with some residual nebular emission.  The
  nebular emission is blue-shifted, behaving the same way as the
  Balmer lines from the nebula in front of the star (Figure
  \ref{fg:he2-36.balmer}, see discussion below).  We conclude that the
  features at $\lambda$4686 and $\lambda$5876 are helium lines from
  the hot star, and that we have confirmed the presence of the hot
  companion of He2-36 in the optical spectrum with the present
  observations.
 
  We used the Atomic Line List
  v2.04\footnote{http://www.pa.uky.edu/~peter/atomic/} to determine if
  there are other helium lines that should have been detected in our
  spectral range of 4120-6730\AA.  The $\lambda$5876 absorption line
  is the strongest \ion{He}{1} line expected in this range.  Of the
  other, fainter \ion{He}{1} lines, $\lambda$6678 shows no absorption
  but with larger uncertainty because this line lies near the edge of
  an order, and $\lambda$5016 and $\lambda$5048 fall in the CCD gap,
  while $\lambda$4471, $\lambda$4388, and $\lambda$4713 coincide with
  the cool star absorption lines, thus are not observable.
  Furthermore, $\lambda$4686 is the strongest expected \ion{He}{2}
  absorption line in our spectrum, the weaker $\lambda$6560 is faintly
  visible in the wings of the H$\alpha$ absorption (Figure
  \ref{fg:he2-36.balmer}, bottom right).

  The spectrum of He~2-36 also shows a feature at $\lambda$4780. Since
  the absorption line looks split in the middle, we focused our
  attention to it, in order to determine its nature.  First, we
  checked against the possibility of nebular emission contamination.
  We extracted the spectrum using a narrower aperture size of 4
  pixels\footnote{Our regular extractions were of 8 pixels.} in order
  to exclude most of the nebular emission.  In Figure
  \ref{fg:he2-36.balmer} we show the 8- and 4-pixel extractions of our
  spectrum around the Balmer lines.  We see from the Figure that the
  8-pixel extraction discloses the broad stellar absorption and the
  narrow nebular emission, while the spectrum of the 4-pixel
  extraction shows only the broad stellar absorption.  For the nebular
  emission, H$\delta$, H$\gamma$, and H$\beta$ (Figure
  \ref{fg:he2-36.balmer}, top three left panels) are blue-shifted
  compared to the central wavelengths of the stellar absorption,
  indicating the emission is mostly from the foreground nebula.  At
  H$\alpha$, where the S/N is the highest, the nebular emission line
  shows a split with a blue-shifted and a red-shifted component
  (Figure \ref{fg:he2-36.balmer}, bottom left panel), indicating that
  the observed emission originates both from the nebula in front and
  behind the central star.  The nebular emission line of \ion{O}{3}
  $\lambda5007$ shows the same structure.  By comparing the spectra
  extracted with the two apertures, it is evident that the narrow
  aperture extraction has excluded most of, if not all, the nebular
  emission.  We compared the $\lambda$4780 feature as it appears with
  the two aperture extractions, and found that the split feature
  appears in both spectra.  Therefore we conclude that this feature
  must be of stellar origin.  It is worth noting that to produce the
  spectra section of Figure \ref{fg:he2-36.balmer} (and all other
  4-pixel extraction) we did not perform any nebular subtraction, but
  simply select the data within the pixels closer to the star.

  Second, we need to establish whether the cool or the hot companion
  is responsible for the split feature $\lambda$4780. If absorbed by
  the cool component, this feature should be interpreted as
  \ion{Ti}{2} $\lambda$4779.99. In this case, though, this would be
  the only split line among the several \ion{Ti}{2} absorption
  observed.  If, on the other hand, this transition originates in the
  hot stellar photosphere, it could be identified as \ion{O}{4}
  $\lambda$4779.10. But even this possibility is puzzling, since in
  the present spectrum we could not detect the other components
  corresponding to the 2s2p(3P0)3p - 2s2p(3P0)3d transition of
  \ion{O}{4}.  If the feature was indeed \ion{O}{4}, and the observed
  splitting was produced by the magnetic field, we could use equation
  (\ref{eq:B}) to estimate the field strength.  This \ion{O}{4}
  transition has $g_{\rm eff}$=4/3, thus the observed 0.7~\AA~split
  could imply a field strength of the order of $\sim$ 25,000 G if we
  assume that the line splits into a doublet in the weak field regime.
  If we use the sequence of spectra of hot sub-dwarfs published by
  \cite{otoole05}, our deduced relative strengths of \ion{He}{2}
  $\lambda$4686 and \ion{He}{1} $\lambda$5876 would appear to indicate
  a star with $\log g\sim 4.5 - 5.0$ and $T_{\rm eff} \sim 50,000 -
  70,000$ K.  Such a star would evolve into a magnetic white dwarf
  with a surface dipole field strength of $\sim 8\times 10^7 -
  2.5\times 10^8$ G if we adopt a $\log g\sim 8.5$ as being typical of
  magnetic white dwarfs.  This field is comfortably within the range
  observed for the field distribution of isolated magnetic white
  dwarfs.

  In Table \ref{tb:split} we list the measurements from the
  $\lambda$4780 feature in He~2-36.  We measured the line as a whole
  to obtain the central wavelength, and fit the split feature with two
  line profiles in order to get the wavelength displacements, both in
  the 8-pixel and the 4-pixel extractions.  The displacements of the
  wavelengths from the central wavelength are the same for short-ward
  and long-ward directions, with differences within the errors
  ($\sim$0.2\AA).  Their fluxes are equal in the 4-pixel measurement,
  but not in the 8-pixel measurement.  This may be because there is a
  slight nebular contamination, but the split feature is not severely
  affected by it. The hot component interpretation of the
  $\lambda$4780 feature needs verification in order to recover the
  other components of the transition multiplet.  Nonetheless, the fact
  that the feature is split into two components of equal strength and
  displaced by the same amount from the zero field position would be
  in agreement with a Zeeman split feature in the low field regime.

  We realize that our proposal that we are seeing absorption lines
  from the hot star in He2-36 superposed on the A star spectrum is not
  without difficulty, since such absorption features would require a
  comparable contribution by the two stars in the optical. In fact, if
  we assume that these two stars are associated and therefore at the
  same distance, and we allow for differences in temperatures and
  gravities, we would only expect approximately a 2\% contribution to
  the optical spectrum by the hot star. A possible resolution to this
  dilemma is that the A star is not physically linked to the nebula
  and the hot component that we have found is indeed the central
  ionizing star of the PN.

  Whether or not we have measured a magnetic field in He2-36 hinges on
  the identification of the split absorption feature at 4780 which we
  have tentatively attributed to \ion{O}{4} from the hot star. This
  remains to be confirmed through the detection of other components of
  the multiplet with their associated Zeeman splitting. Higher
  resolution data should also reveal Zeeman splitting in the helium
  lines.

\section{Conclusions}
\label{sec:dis}

  We have investigated the shaping mechanism of bipolar PNe and the
  progenitors of MWDs using high resolution echelle spectroscopy of
  central stars of eight southern bipolar PNe.  We looked for Zeeman
  splitting of the stellar lines caused by possible magnetic fields
  associated with the CSs.  Among the eight objects, we did not detect
  any stellar emission or absorption lines for He~2-25, NGC~2818, and
  He~2-186.  HDW~5 have one absorption and He~2-123 have two faint
  \ion{C}{4} emission lines detected.  The spectra of He~2-36,
  He~2-64, and MyCn~18 have much higher S/N and various stellar lines
  in their spectra were detected.  For those with stellar features, we
  have assigned a spectral class mainly following
  \citeauthor{mendez91}'s \citeyearpar{mendez91} classification.  An
  expanding atmosphere has also been found for the central star of He
  2-64 with a velocity of $\sim$ 100 km~s$^{-1}$ as derived from the
  P-Cygni profiles of the \ion{He}{1} lines.

  We have placed upper limits of $\sim 20,000$~G for the CSs of
  He~2-64, and MyCn~18. We could not perform our search for Zeeman
  splitting in HDW~5, He~2-25, NGC~2818, He~2-123, and He~2-186 due to
  limited S/N in their spectra, and thus no upper limits to the
  magnetic fields associated with the nuclei of those PNe could be
  placed. The low detection rate of stellar features in the central
  stars indicate that even though we have chosen the brightest objects
  possible, the S/N from the observations with a 4-m telescope are
  still too low for our science goals.  In order to collect a proper
  sample for our analysis, it is essential to obtain high sensitivity
  observations of the central star spectra with larger aperture
  telescopes.

  We found \ion{He}{1} and \ion{He}{2} absorption lines in the
  spectrum of the CS of He~2-36, disclosing the hot component of the
  possibly binary CS.  The split feature at $\lambda$4780 in He~2-36
  is intriguing.  If the feature is from the hot companion of the
  system, and the split is really caused by Zeeman splitting, this
  would be the first direct detection in the flux spectrum of a
  magnetic field in a CS of a PN.  Further observations with larger
  aperture telescopes and conducted at different epochs will be
  crucial to rule out the possibility that the split is caused by
  binarity.  In addition, we plan to obtain spectropolarimetry to
  confirm the presence of a magnetic field in the central star of
  He~2-36.  This is a complicated yet very exciting stellar system and
  requires more observations to further understand it.

  Our limits to the magnetic field in the sample stars indicate that,
  if required to produce the bipolar shape, magnetic fields do not
  need to be stronger than a few tens of kilogauss, in broad agreement
  with the Garcia-Segura models. Our results are supplementary to
  those of \cite{jordan05} where possible fields of a few kilo-Gauss
  were reported from spectropolarimetric studies of the CSs of 4
  asymmetric PNe.

  \acknowledgments

  Many thanks to Olivier Hainaut, Emanuela Pompei, Ivo Saviane, and
  Jeremy Walsh for their help with the observation and data reduction,
  and to Jason Aufdenberg, Mike Barlow, Guillermo Garcia-Segura and
  Bill Sherry for scientific discussion.

\clearpage

\begin{figure}
\plotone{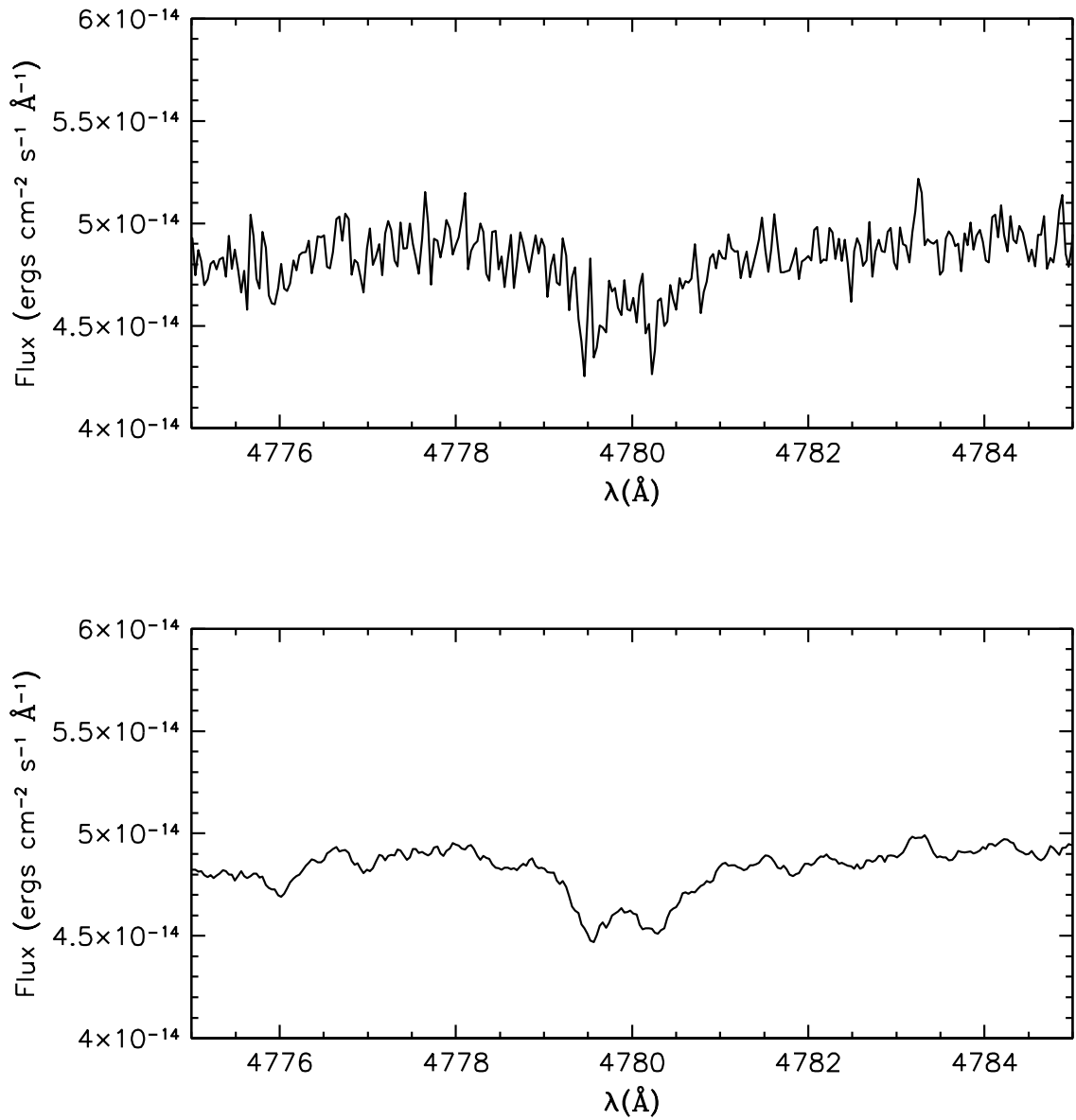}
\caption{Spectrum of He~2-36.  $\lambda$4780 absorption line showing
  splitting into two components.  The bottom panel is the smoothed
  spectrum, which shows the splitting more clearly.}
\label{fg:he2-36.4780}
\end{figure}

\begin{figure}
\includegraphics[angle=270, width=80mm]{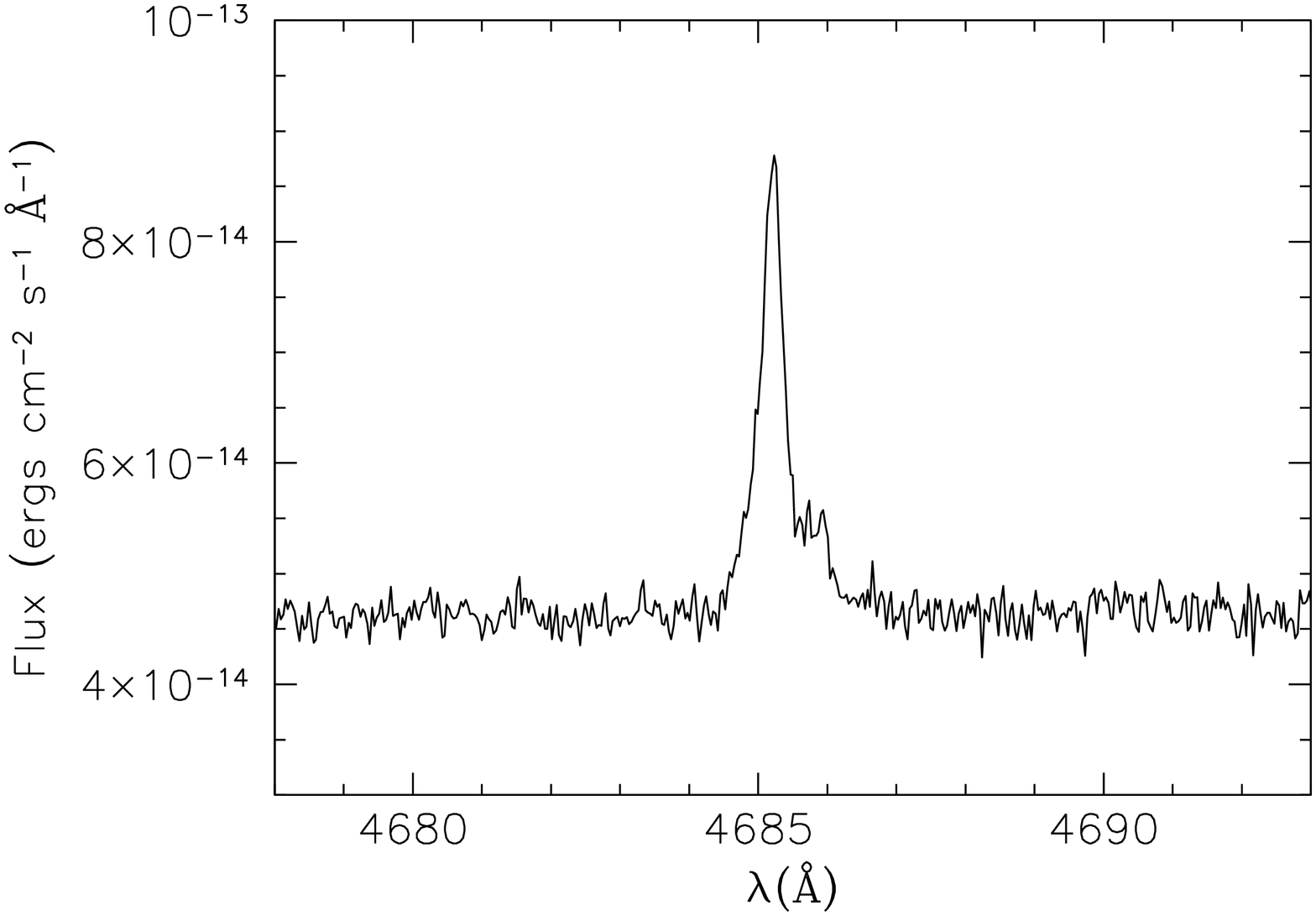}
\includegraphics[angle=270, width=80mm]{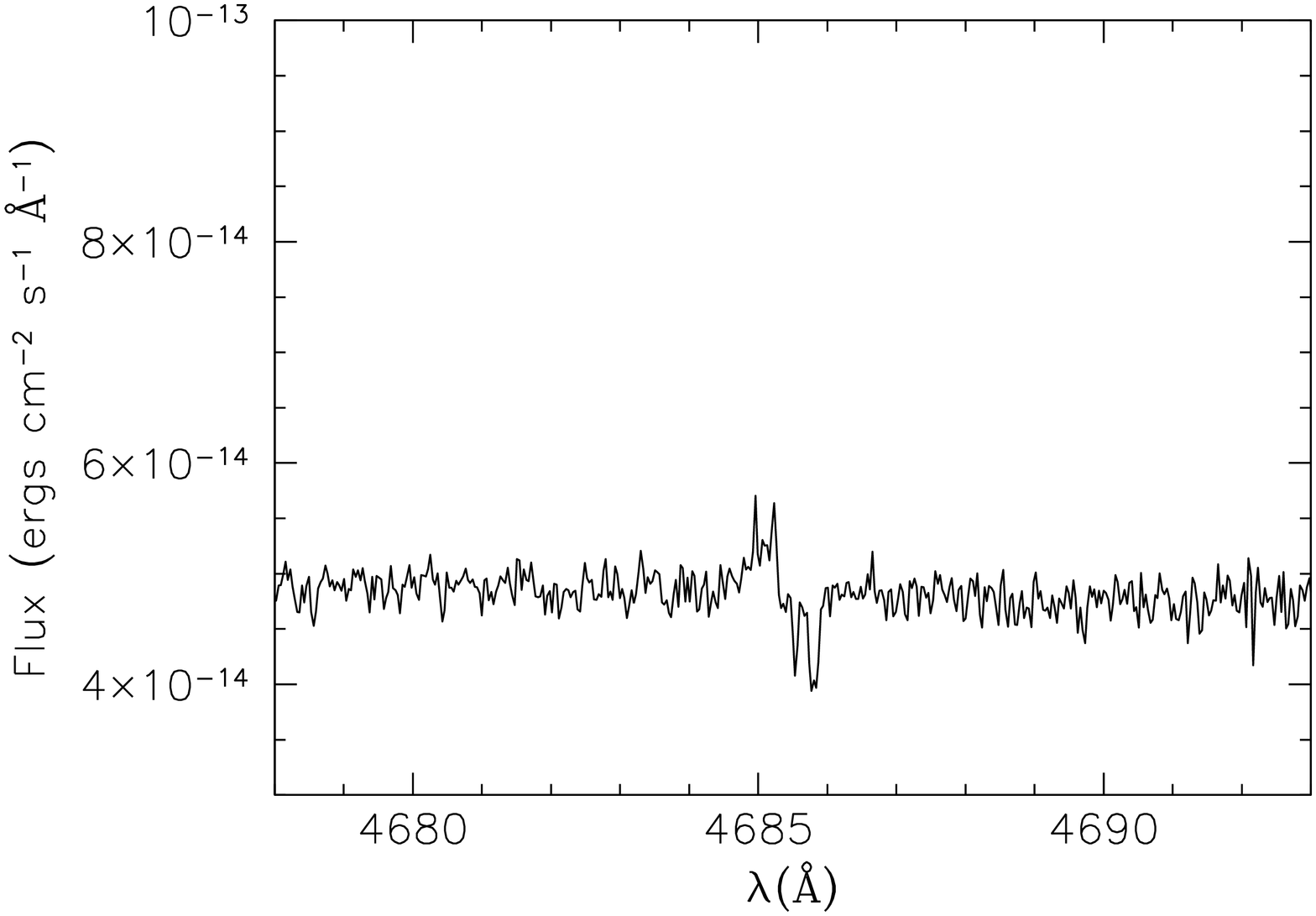}
\caption{Spectrum of He~2-36.  \ion{He}{2} at 4686~\AA~with extraction
  aperture of 8 pixels (left) and 4 pixels (right).  The stellar
  absorption line can be better seen with the narrower aperture.}
\label{fg:he2-36.4686}
\includegraphics[angle=270, width=80mm]{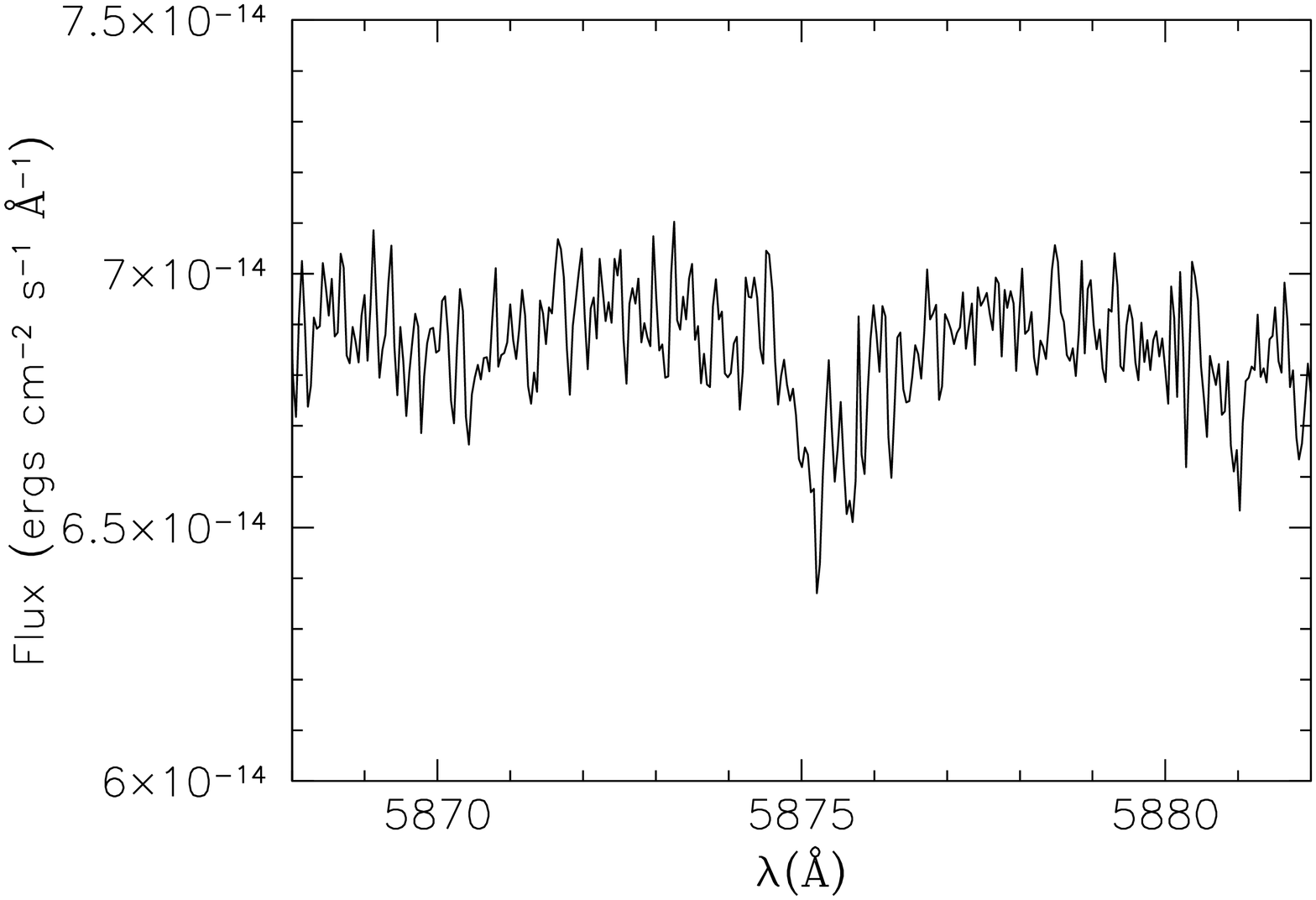}
\caption{Spectrum of He~2-36. \ion{He}{1} at 5876~\AA.}
\label{fg:he2-36.5876}
\end{figure}

\begin{figure}
\includegraphics[angle=270, width=70mm]{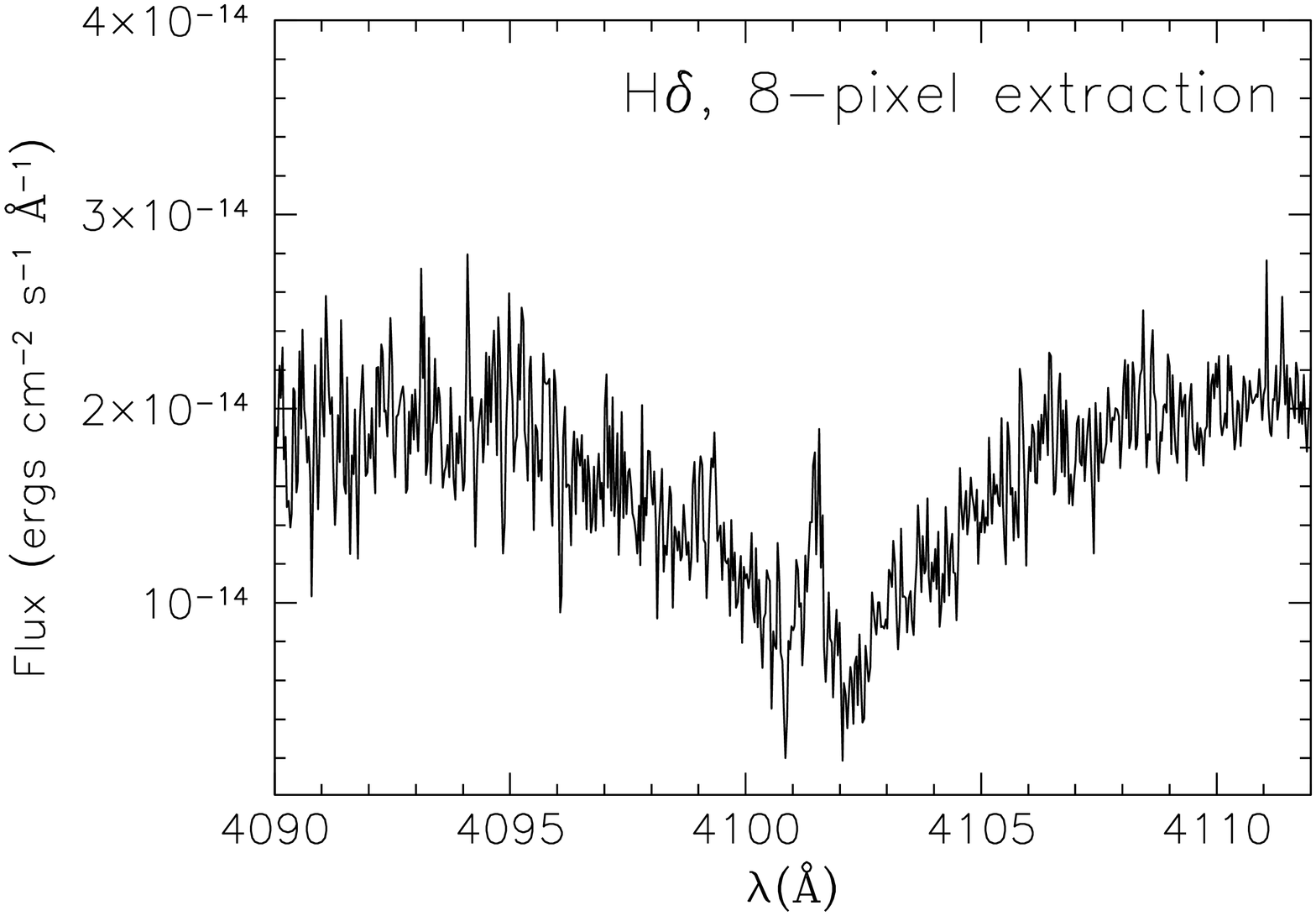}
\includegraphics[angle=270, width=70mm]{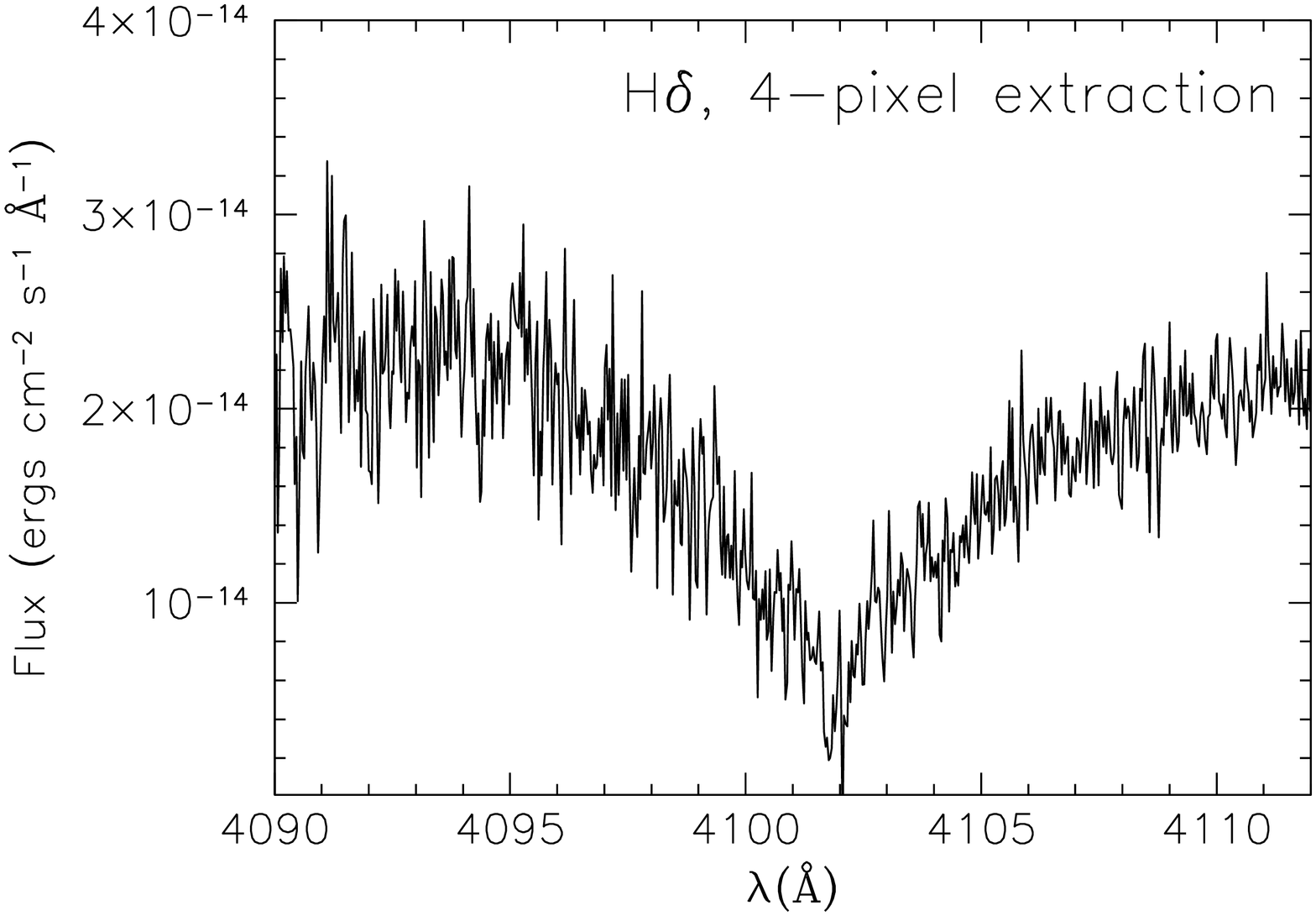} \\
\includegraphics[angle=270, width=70mm]{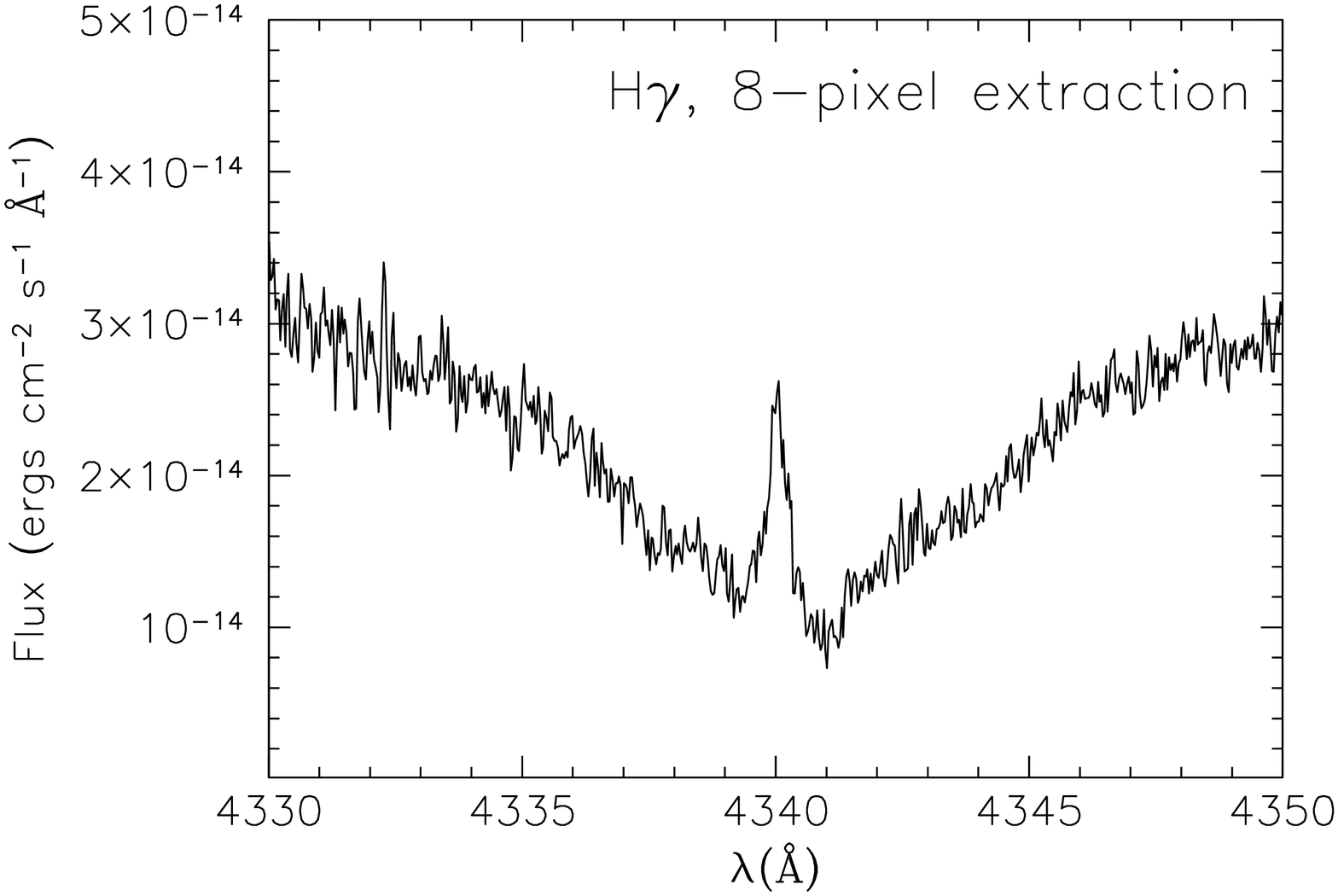}
\includegraphics[angle=270, width=70mm]{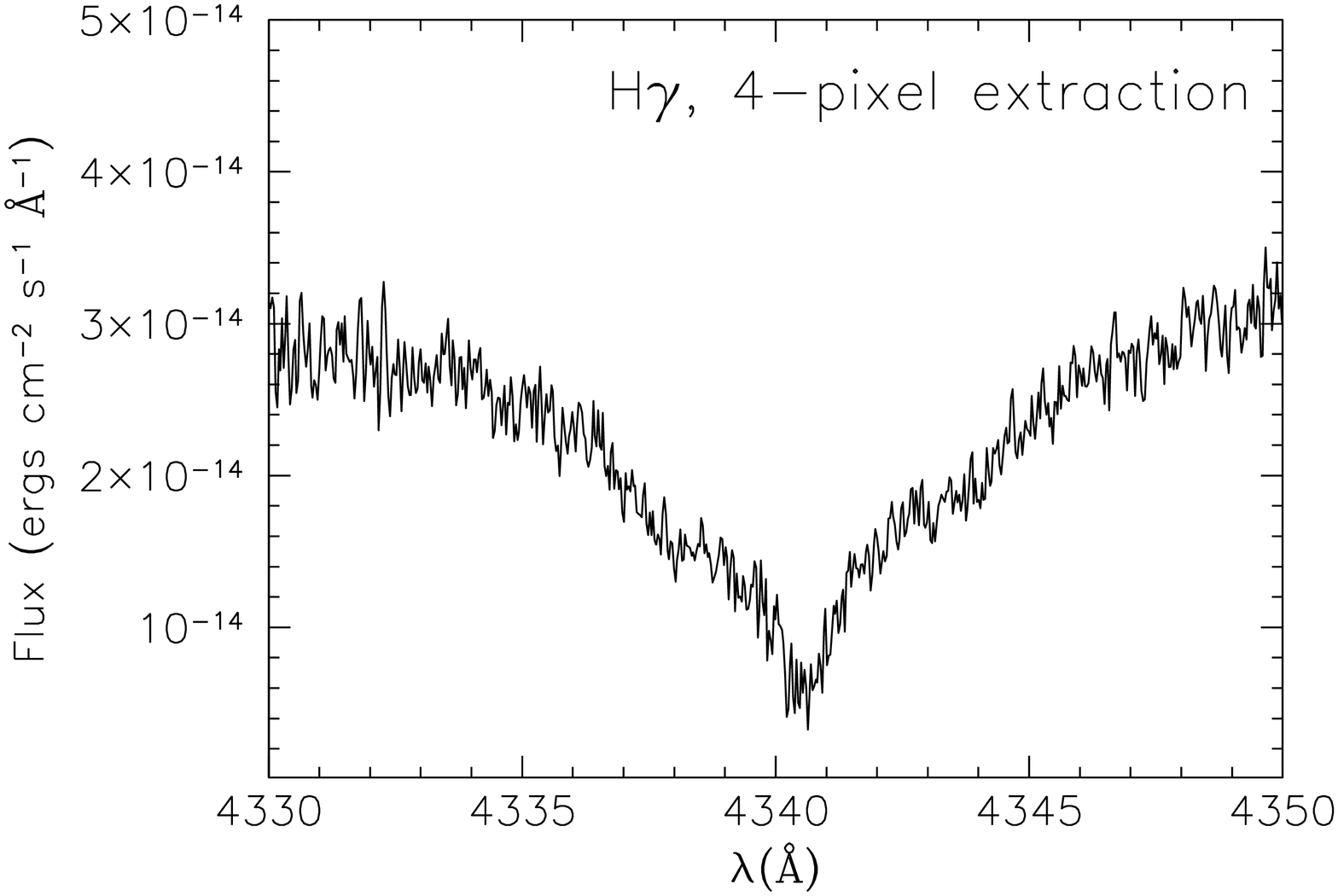} \\
\includegraphics[angle=270, width=70mm]{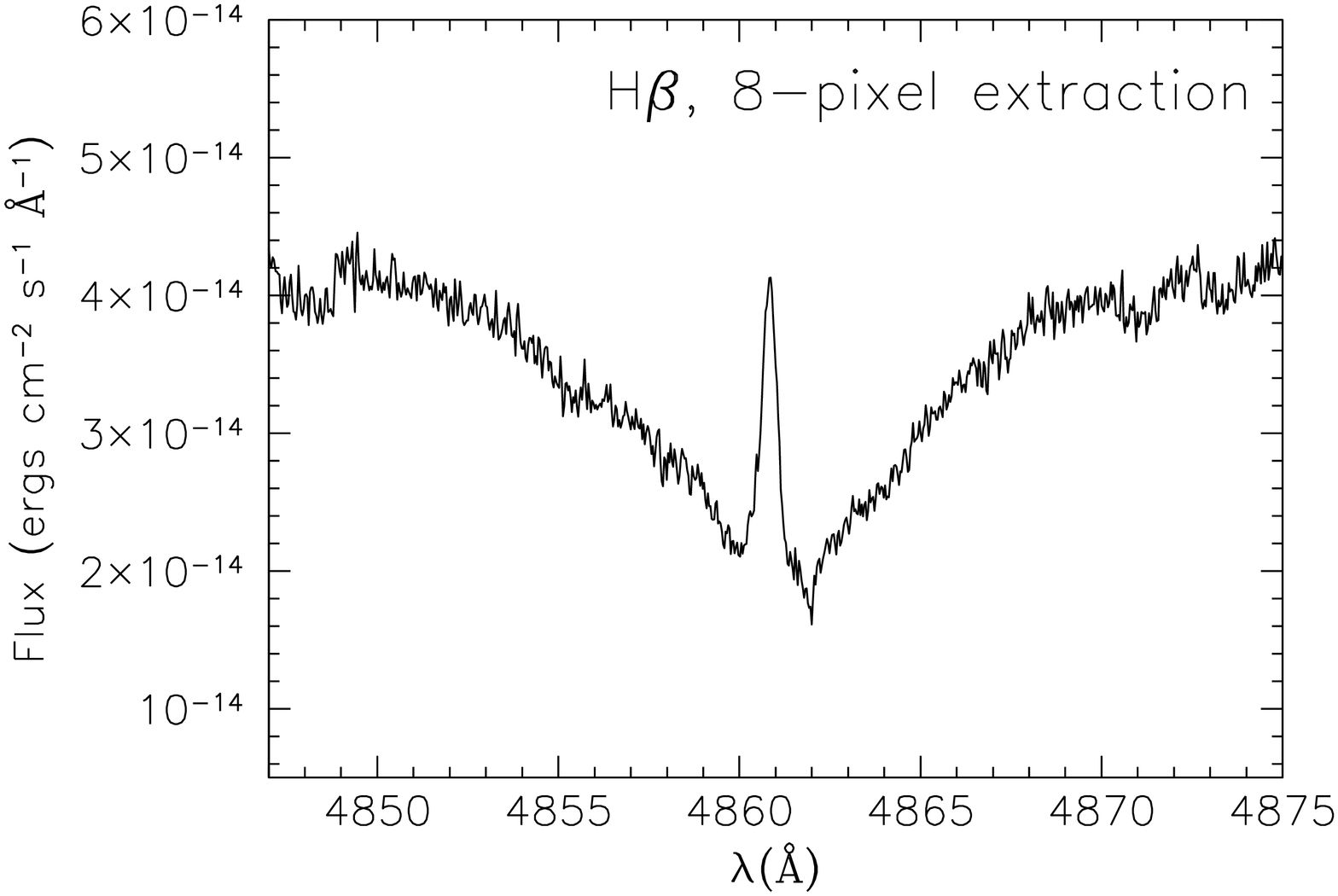}
\includegraphics[angle=270, width=70mm]{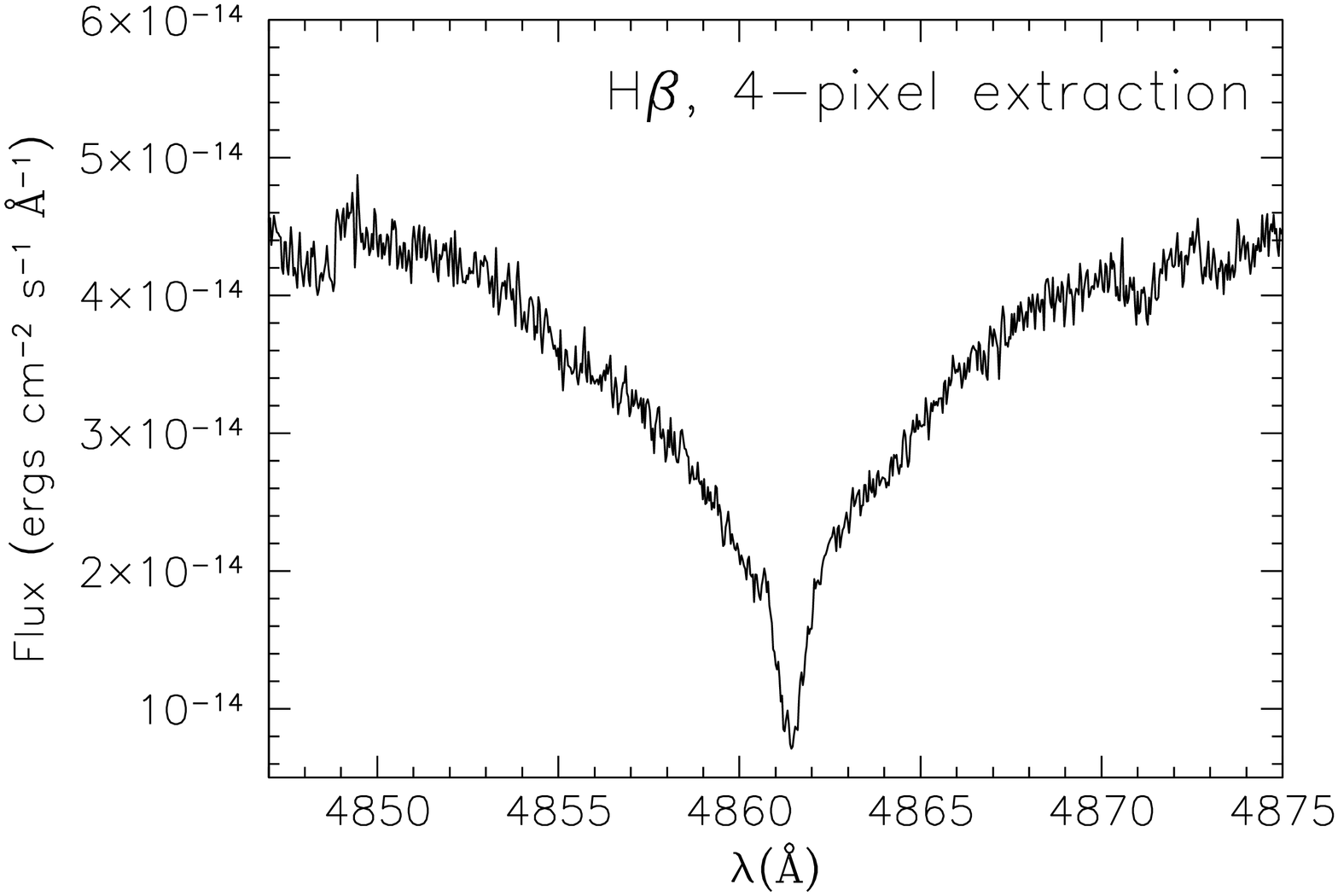} \\
\includegraphics[angle=270, width=70mm]{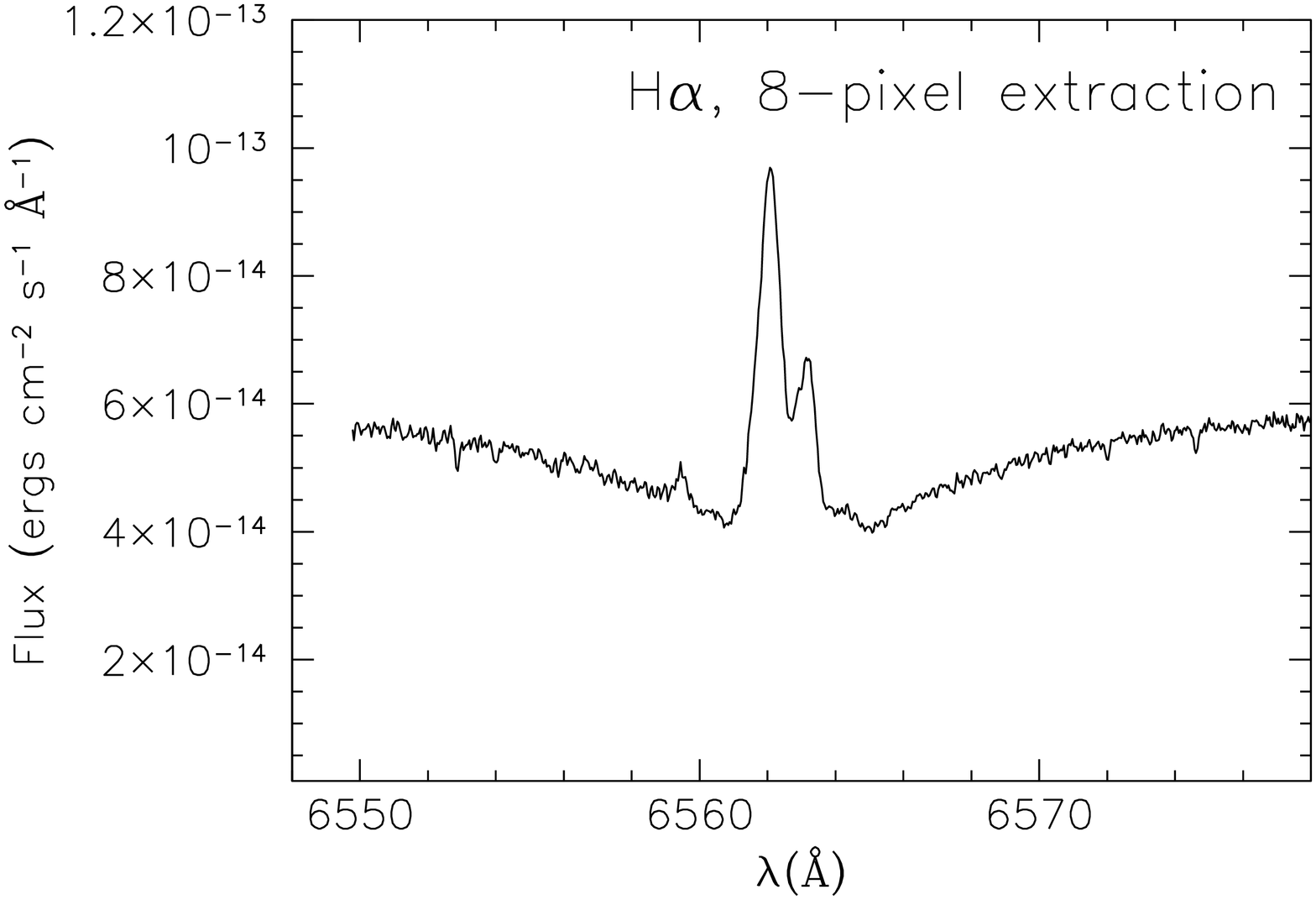}
\includegraphics[angle=270, width=70mm]{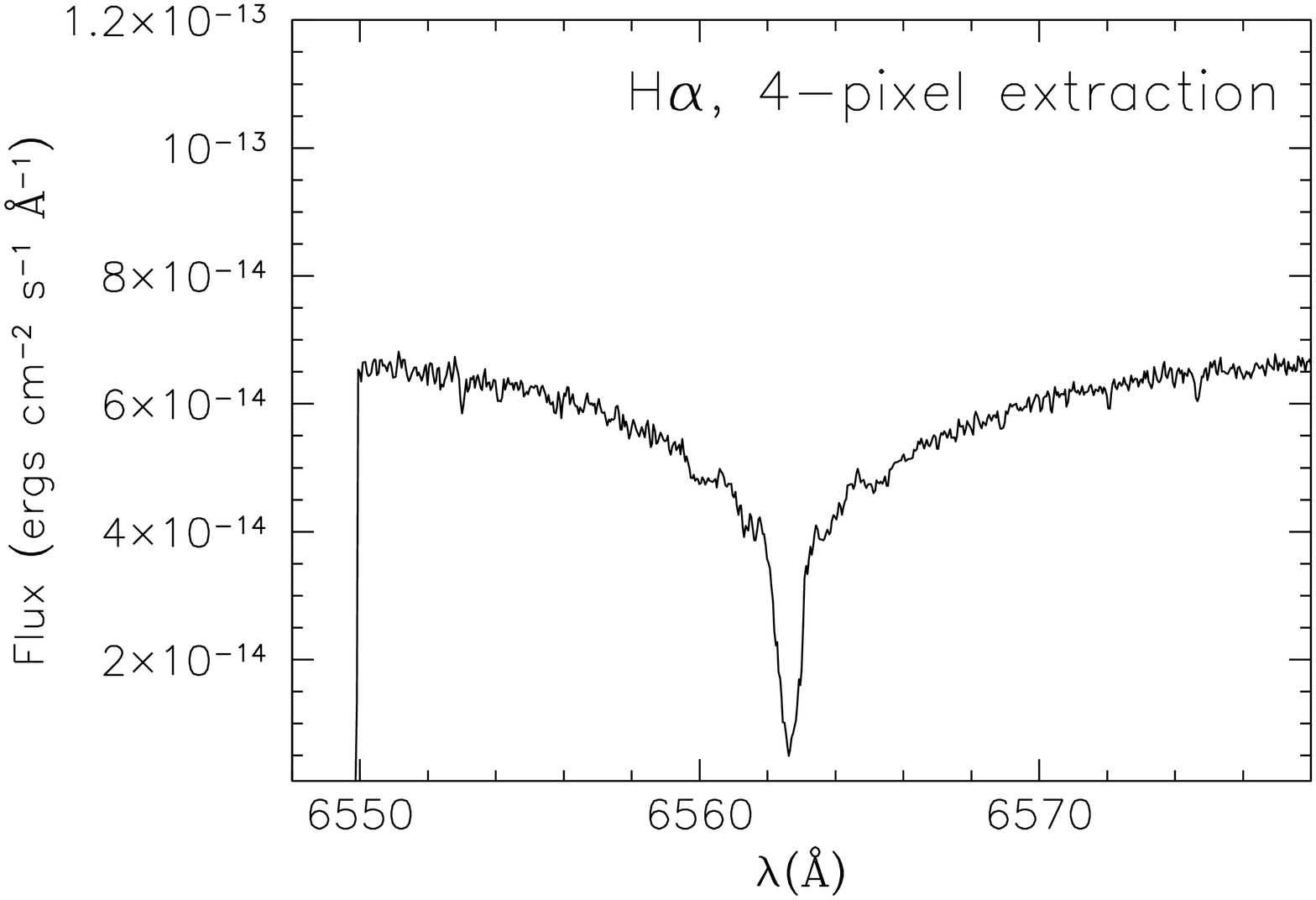}
\caption{Spectrum of He~2-36: H$\delta$ at 4101~\AA~(first row),
  H$\gamma$ at 4340~\AA~(second row), H$\beta$ at 4861~\AA~(third
  row), and H$\alpha$ at 6563~\AA~(bottom row) with extraction
  apertures of 8 pixels (left panels) and 4 pixels (right panels).}
\label{fg:he2-36.balmer}
\end{figure}

\clearpage

\begin{deluxetable}{lcccccc}
\tablecaption{Target List\label{tb:source}}
\tablehead{ 
\colhead{Object} & \colhead{PN G} & \colhead{B} & \colhead{V}
& \colhead{Mag Ref} & \colhead{Tz(He {\footnotesize II})}
& \colhead{Tz Ref} \\
& & & & & \colhead{$\times 10^3$ K}
}
\startdata
HDW 5 & PN G218.9$-$10.7 & 16.55 & 16.29 & 2 & $\dots$& $\dots$ \\
He 2-25 & PN G275.2$-$03.7 & 17.08 & 16.96 & 1 & $<$61 & 4 \\
He 2-36 & PN G279.6$-$03.1 & 11.96 & 11.37 & 1 & $\dots$& $\dots$\\
NGC 2818 & PN G261.9+08.5 & 19.58 & $\dots$& 3 & 215.0 & 4 \\
He 2-64 & PN G291.7+03.7 & $\dots$& $\dots$& $\dots$& 48.3\tablenotemark{*} & 5\\
He 2-123 & PN G323.9+02.4 & 17.55 & 16.84 & 1 & $<$60 & 4 \\
He 2-186 & PN G336.3$-$05.6 &  18.34 & 16.62 & 1 & 95.5 & 4 \\
MyCn 18 & PN G307.5$-$04.9 & 14.5 & $\dots$& 2 & 51.6 & 4 \\
\enddata
\tablerefs{1. \cite{shaw89}; 2. \cite{tylenda91}; 3. \cite{gathier88};
  4. \cite{phillips03}, and references therein;
  5. \cite{preite-martinez89}.}
\tablenotetext{*}{Energy-Balance temperature.}
\end{deluxetable}

\begin{deluxetable}{lccccc}
\tablecaption{Filter Properties\label{tb:filters}}
\tablehead{\colhead{Filter Number} & \colhead{Type} & \colhead{Central
    Wavelength} & \colhead{Bandwidth} & \colhead{Pixel Size}
   \\ & & \colhead{(nm)} & \colhead{(nm)} & \colhead{(arcsec)}}
\startdata
602 & U Bessel & 354.208 & 54.176 & 0.37  \\
603 & B Bessel & 422.305 & 94.0709 & 0.37  \\
606 & V Bessel & 542.605 & 104.471 & 0.33  \\
608 & R Bessel & 640.962 & 154.175 & 0.33  \\
\enddata
\end{deluxetable}

\begin{deluxetable}{lcccccc} 
\tablecaption{Observing Logs\label{tb:ob-log}}
\tablehead{& &\multicolumn{5}{c}{Exposures in seconds} \\ \cline{3-7} 
\colhead{Object} & \colhead{Observation Date} &
\colhead{Echelle} & \colhead{U\#602} & \colhead{B\#603} & \colhead
{V\#606} & \colhead{R\#608}
}
\startdata
HDW 5 & Feb 4, 2003 & 4x1800 & 20x2 & 20x2 & 20 & 20 \\
He 2-25 & Feb 4, 2003 & 4x1800 & 20 & 20 & 20 & 20 \\
He 2-36 & Feb 4, 2003 & 2x900 & 20 & 15 & 5 & 5 \\
NGC 2818 & Feb 5, 2003 & 2x1800 & 20 & 20 & 20 & 20 \\
He 2-64 & Feb 5, 2003 & 4x1800 & 20 & 20 & 20 & 20 \\
He 2-123 & Feb 5, 2003 & 1x1200+3x1800 & 20 & 20 & 20 & 20 \\
He 2-186 & Feb 5, 2003 & 2x1800 & 20 & 20 & 20 & 20 \\
MyCn 18 & Feb 4, 2003 & 4x900 & 20 & 20 & 20+5 & 20+5 \\
\enddata
\end{deluxetable}

\begin{deluxetable}{lccccccc}
\tablecaption{Nebular Line Intensities\label{tb:line_ratios}} \tablehead{
\colhead{Object} & \colhead{$F_{\rm H\beta}$} & \colhead{4686\AA} &
\colhead{4958\AA} & \colhead{5007\AA} & \colhead{6563\AA} &
\colhead{6584\AA} & \colhead{\it c} \\ &
\colhead{[ergs~cm$^{-2}$~s$^{-1}$~\AA$^{-1}$]} & \colhead{\ion{He}{2}} &
\colhead{[\ion{O}{3}]} & \colhead{[\ion{O}{3}]}
& \colhead{H$\alpha$} & \colhead{[\ion{N}{2}]} & 
}
\startdata 
He 2-25 & 1.90e-13\tablenotemark{a} & 0.04 & 145\tablenotemark{a} &
579\tablenotemark{a} & 2381\tablenotemark{a} & 17\tablenotemark{a} &
2.63 \\
He 2-36 & 1.24e-14\tablenotemark{a} & 196\tablenotemark{a} &
385\tablenotemark{a} & 1502\tablenotemark{a} & 545\tablenotemark{a} &
\nodata & 0.80 \\
NGC 2818 & 5.34e-15\tablenotemark{a} & 75\tablenotemark{a} &
390\tablenotemark{a} & 1249\tablenotemark{a} & 549\tablenotemark{a} &
584\tablenotemark{a} & 0.81 \\
He 2-64 & 2.63e-14 & Stellar abs. & 17\tablenotemark{a} &
91\tablenotemark{a} & 548 & 264\tablenotemark{a} & 0.81 \\
He 2-123 & 1.05e-13 & \nodata & 73\tablenotemark{a} &
239\tablenotemark{a} & 1495 & 852\tablenotemark{a} & 2.06 \\
He 2-186 & 1.97e-13\tablenotemark{a} & 47 & 549 &
1852\tablenotemark{a} & 841\tablenotemark{a} & 516\tablenotemark{a} &
1.34 \\
MyCn 18 & 1.83e-12 & Stellar emi. & 113 & 290 & 590 &
282\tablenotemark{b} & 0.90 \\
\enddata 
\tablenotetext{a}{Integrated over the area} 
\tablenotetext{b}{Wide wings}
\end{deluxetable}

\begin{deluxetable}{ccccc}
\tablecaption{Stellar Line Analysis\label{tb:sline}} 
\tablehead{
\colhead{$\lambda$} & \colhead{ID} & \colhead{$\lambda_{\rm obs}$} & 
\colhead{Flux} & \colhead{W} \\
\colhead{[\AA]} & & \colhead{[\AA]} &
\colhead{[ergs~cm$^{-2}$~s$^{-1}$~\AA$^{-1}$]} & \colhead{[\AA]} 
} 
\startdata
HDW 5 \\
 & N. I. & 5757.1 & -6.58e-16 & 0.7198  \\
\\
He2-36 \\
4101 & H$\delta$ & 4101.7 & -7.9e-14 & 4.0 \\
4340 & H$\gamma$ & 4340.4 & -1.8e-13 & 5.8 \\
4861 & H$\beta$ & 4861.0 & -2.1e-13 & 4.9 \\
4686 & \ion{He}{2} & 4685.7 & -2.6e-15 & 0.05 \\
 & N. I. & 5780.1 & -1.9e-14 & 0.3 \\
5877 & \ion{He}{1} & 5875.4 & -3.7e-15 & 0.05 \\
6563 & H$\alpha$ & 6563.0 & -1.8e-13 & 3.1 \\
\\
He2-64 \\
4686 & \ion{He}{2} & 4684.8  & -7.5e-16 & 0.4   \\
 & N. I. & 5281.8 & 2.1e-14 & -11.8 \\
5411 & \ion{He}{2} & 5411.0 & -1.4e-15 & 0.7 \\
 & N. I. & 5759.0 & -2.0-15 & 1.1 \\
\\
He 2-123 \\
5801 & \ion{C}{4} & 5799.6 & 1.1e-15 & -1.7 \\
5811 & \ion{C}{4} & 5810.7 & 8.3e-16 & -1.3 \\
\\
MyCn 18 \\
4541 & \ion{He}{2} & 4539.7 & -6.5e-15 & 0.8 \\
4634 & \ion{N}{3} & 4632.6 & 5.0e-15 & -0.6 \\
4658 & \ion{C}{4} & 4656.2 & 9.3e-15 & -1.1 \\
4686 & \ion{He}{2} & 4684.7 & 2.0e-14 & -2.5 \\
 & N. I. & 5779.7 & -1.5e-15 & 0.2  \\
5811 & \ion{C}{4} & 5810.2 & 4.2e-15 & -0.6  \\
\enddata
\end{deluxetable}

\begin{deluxetable}{cccccc}
\tablecaption{P-Cygni profiles in He~2-64\label{tb:p-cygni}} 
\tablehead{
\colhead{$\lambda$} & \colhead{ID} & \colhead{$\lambda_{\rm obs}$} & 
\colhead{Flux} & \colhead{W} \\
\colhead{[\AA]} & & \colhead{[\AA]} &
\colhead{[ergs~cm$^{-2}$~s$^{-1}$~\AA$^{-1}$]} & \colhead{[\AA]}
} 
\startdata
4471 & \ion{He}{1} & 4471.8 & -4.0e-16 & 0.3 \\
& & 4471.5 & 5.3e-16 & -0.3 \\
5876 & \ion{He}{1} & 5874.6 & -6.5e-16 & 0.4 \\
& & 5875.7 & 2.9e-15 & -1.9 \\
6678 & \ion{He}{1} & 6679.4 & -5.0e-16 & 0.5 \\
& & 6680.7 & 1.1e-15 & -1.1 \\
\enddata
\end{deluxetable}
 
\begin{deluxetable}{ccc}
\tablecaption{ He~2-36 at 
  $\lambda$4780 \label{tb:split}}
\tablehead{
\colhead{$\lambda_{\rm obs}$} & \colhead{Flux} & \colhead{W} \\
\colhead{[\AA]} & \colhead{[ergs~cm$^{-2}$~s$^{-1}$~\AA$^{-1}$]} &
  \colhead{[\AA]}  
}
\startdata
\multicolumn{3}{l}{Single line measurement} \\ 
  4779.9 & -5.2e-15 & 0.1 \\ \\
\multicolumn{3}{l}{Deblended lines, 8-pixel aperture} \\
  4779.5 & -1.5e-15 & 0.03 \\
  4780.2 & -2.5e-15 & 0.05 \\ \\
\multicolumn{3}{l}{Deblended lines, 4-pixel aperture} \\
  4779.6 & -1.7e-15 & 0.03 \\
  4780.2 & -1.8e-15 & 0.04 \\
\enddata
\end{deluxetable}

\end{document}